\documentclass{article}

 \usepackage[main, final]{neurips_2026}

\usepackage[utf8]{inputenc} 
\usepackage[T1]{fontenc}    
\usepackage{hyperref}       
\usepackage{url}            
\usepackage{booktabs}       
\usepackage{amsfonts}       
\usepackage{nicefrac}       
\usepackage{microtype}      
\usepackage{xcolor}         
\usepackage{graphicx}
\usepackage{subcaption}
\usepackage{amsmath} 
\usepackage{colortbl}
\usepackage{cleveref}
\usepackage[most]{tcolorbox}
\usepackage{xspace}
\usepackage{wrapfig}
\usepackage{algorithm}
\usepackage{algpseudocode}

\definecolor{back_color}{gray}{0.95} 
\definecolor{easy_color}{HTML}{2e4057}
\definecolor{medium_color}{HTML}{083d77}
\definecolor{hard_color}{HTML}{DA4167}
\definecolor{traj_color}{HTML}{3092BF}
\definecolor{oursrow}{RGB}{233,243,253}   
\definecolor{oraclerow}{RGB}{255,243,230} 
\definecolor{headergray}{gray}{0.94}

\newcommand{\method}{UniPath\xspace}

\title{UniPath: Adaptive Coordination of Understanding and Generation for Unified Multimodal Reasoning}

\author{%
  Hayes Bai$^{1}$ \quad Yinyi Luo$^{2}$ \quad Wenwen Wang$^{2}$ \quad
  Qingsong Wen$^{3}$ \quad Jindong Wang$^{1}$\thanks{Corresponding author: \texttt{jdw@wm.edu}.} \\
  \normalfont $^{1}$William \& Mary \quad $^{2}$Carnegie Mellon University \quad
  \normalfont $^{3}$Squirrel Ai Learning
}

\hypersetup{
  pdftitle={UniPath: Adaptive Coordination of Understanding and Generation for Unified Multimodal Reasoning},
  pdfauthor={Hayes Bai, Yinyi Luo, Wenwen Wang, Qingsong Wen, Jindong Wang}
}

\begin{document}

\maketitle

\begin{abstract}
Unified multimodal models (UMMs) aim to integrate understanding and generation within a single architecture. 
However, it remains underexplored how to effectively coordinate these two capabilities for more effective and efficient reasoning. 
Existing coordination approaches either perform coupling during training, without explicit inference-time coordination, or impose a fixed coordination pattern for all inputs.
In this work, we show that multimodal tasks exhibit substantial coordination-path diversity: different inputs favor different coordination paths. 
This suggests that exploiting such diversity is key to improving performance.
We propose \textbf{UniPath}, a framework for adaptively modeling and exploiting coordination-path diversity. 
Instead of enforcing a single coordination pattern, we represent task solving as the selection and execution of a path, ranging from direct answering to textual inference, visual-thought construction, and hypothesis-based exploration. 
We construct role-aligned trajectories to train a path-conditioned executor and introduce a lightweight planner mechanism to enable input-dependent path selection.
Experiments show that leveraging coordination-path diversity improves performance over fixed coordination strategies while providing interpretable intermediate behaviors. The code is available at: \url{https://github.com/AIFrontierLab/TorchUMM}.
\end{abstract}

\section{Introduction}

Unified multimodal models (UMMs) are a new family of models that can perform both understanding and generation tasks within a single architecture. Recent models have shown strong results on visual question answering and image generation~\citep{wang2024emu3, team2023gemini, bai2025qwen3, wu2024janus, chen2025janus, ma2024janusflow}, suggesting that one single model can possess both capabilities. A natural next step is to move from capability coexistence toward effective \textit{coordination:} understanding should provide useful evidence for generation, and generation-side visual signals should in turn support subsequent reasoning.

Coordination affects more than accuracy. If the model chooses a suitable reasoning path, it can use deeper multimodal steps only when useful, reduce unnecessary output tokens, and provide a readable explanation of why a particular solving strategy was used. Poor coordination has the opposite effect: simple paths may overlook problems that need intermediate reasoning, while forcing every input through a long coordination pattern wastes computation and is more prone to errors.

Existing work has explored coordination from different angles. Some methods promote coordination by coupling understanding and generation during training, such as self-play or reconstruction alignment~\citep{su2025unigame, xie2025reconstruction}, improving consistency between perception and synthesis. However, they usually do not explicitly specify when and how coordination should occur at inference time, which limits how much learned cooperation can be exploited. Other methods introduce intermediate textual or visual representations~\citep{qin2025uni}, or use explicit coordination patterns such as analyzing-drafting loops and interleaved reasoning-generation traces~\citep{adloop, irg}. They make coordination more visible, but the protocol is usually fixed during training and inference that do not sufficiently account for the properties of different tasks and questions, making coordination less flexible than needed.

\textit{Do different inputs actually benefit from different coordination strategies?} To answer it, we evaluate BAGEL~\citep{deng2025emerging} under several paths, including direct answering, explicit understanding, textual reasoning, visual-thought construction, and hypothesis exploration (formalized in \S\ref{sec:problem}). \Cref{fig:intro_motivation} illustrates these paths with representative examples: simple perception questions may be answered after understanding alone, while others benefit from textual reasoning, visual-thought construction, or hypothesis exploration.
We then examine whether such path differences translate into measurable performance variation on MMMU~\mbox{\citep{yue2024mmmu}}, a multidisciplinary benchmark spanning expert-level questions across diverse subjects. At the subject level, \Cref{fig:intro_subject_affinity} shows that no single path consistently dominates: different subjects favor different paths, and the best path varies across subjects. At the instance level, \Cref{fig:intro_question_correctness} further shows that correctness varies sharply across paths, with many inputs being solved by only a subset of paths. The complete results for the two heatmaps are in Appendix~\ref{app:path_diversity_full}.
The oracle results shown in \Cref{fig:intro_subject_affinity} provide direct evidence for the value of this diversity. By selecting the best path per input, the oracle substantially outperforms any fixed path, showing that coordination-path diversity is not redundant but can translate into large performance gains.

\newsavebox{\introleftpanel}
\begin{figure*}[t]
  \centering
  \sbox{\introleftpanel}{\begin{minipage}[t]{0.48\linewidth}
    \vspace{0pt}
    \centering
    \begin{subfigure}[t]{\linewidth}
      \centering
      \includegraphics[width=\linewidth]{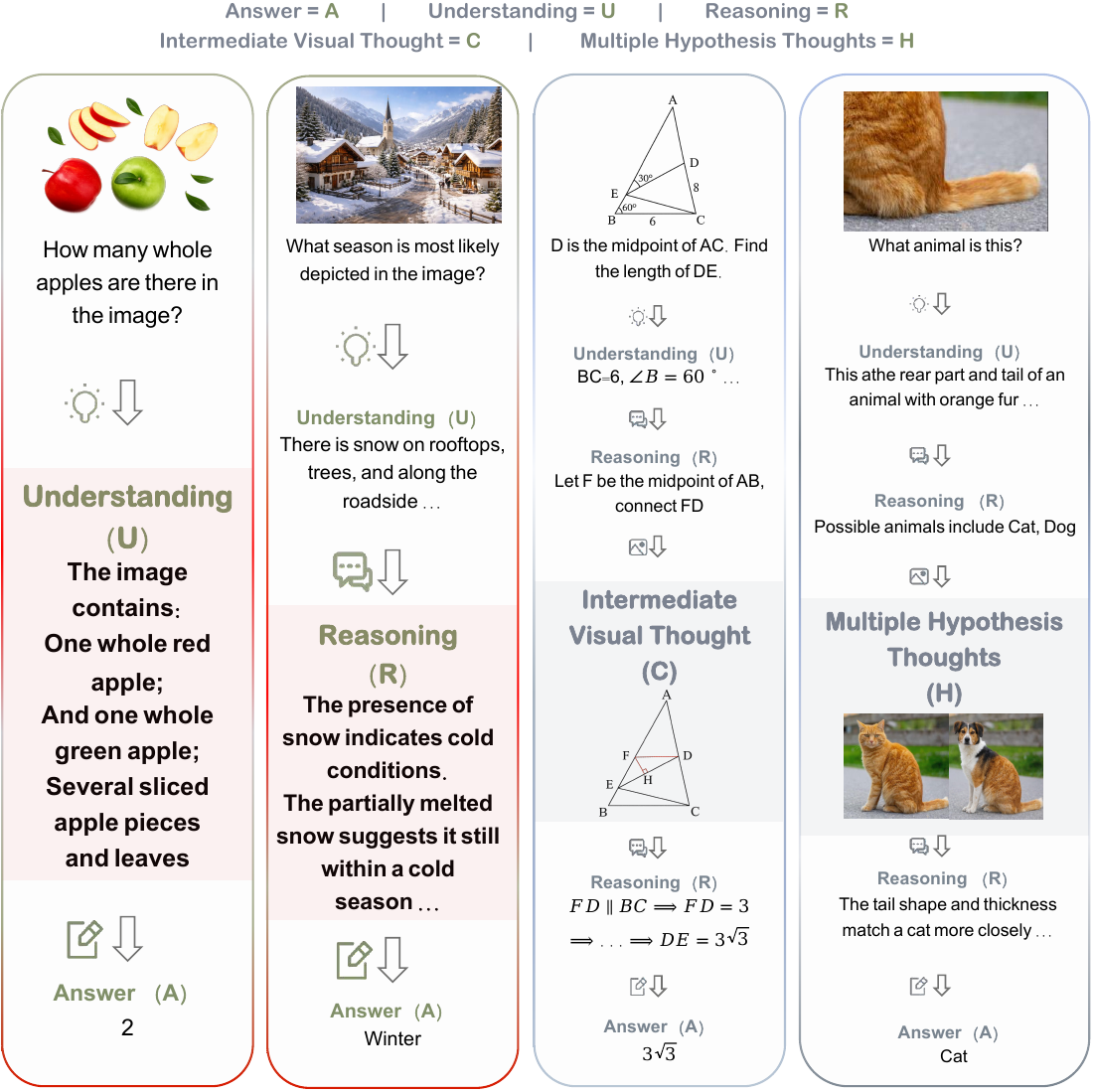}
      \vspace{-.15in}
      \caption{Different inputs favor different coordination paths.}
      \label{fig:intro_path_examples}
    \end{subfigure}
  \end{minipage}}
  \usebox{\introleftpanel}
  \hfill
  \begin{minipage}[t][\dimexpr\ht\introleftpanel+\dp\introleftpanel-1.7\baselineskip\relax][s]{0.50\linewidth}
    \vspace{0pt}
    \centering
    \begin{subfigure}[t]{\linewidth}
      \centering
      \includegraphics[width=\linewidth]{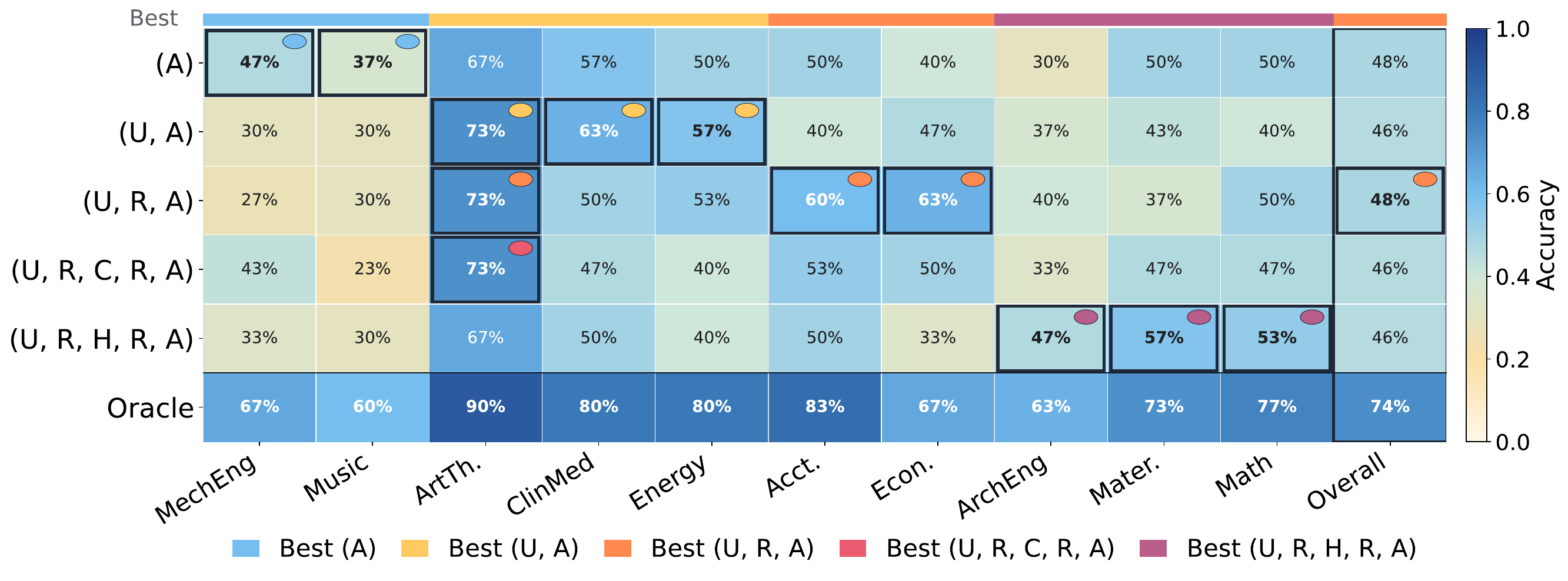}
      \caption{Subject-level path affinity on MMMU.}
      \label{fig:intro_subject_affinity}
    \end{subfigure}
    \vfill
    \begin{subfigure}[t]{\linewidth}
      \centering
      \includegraphics[width=\linewidth]{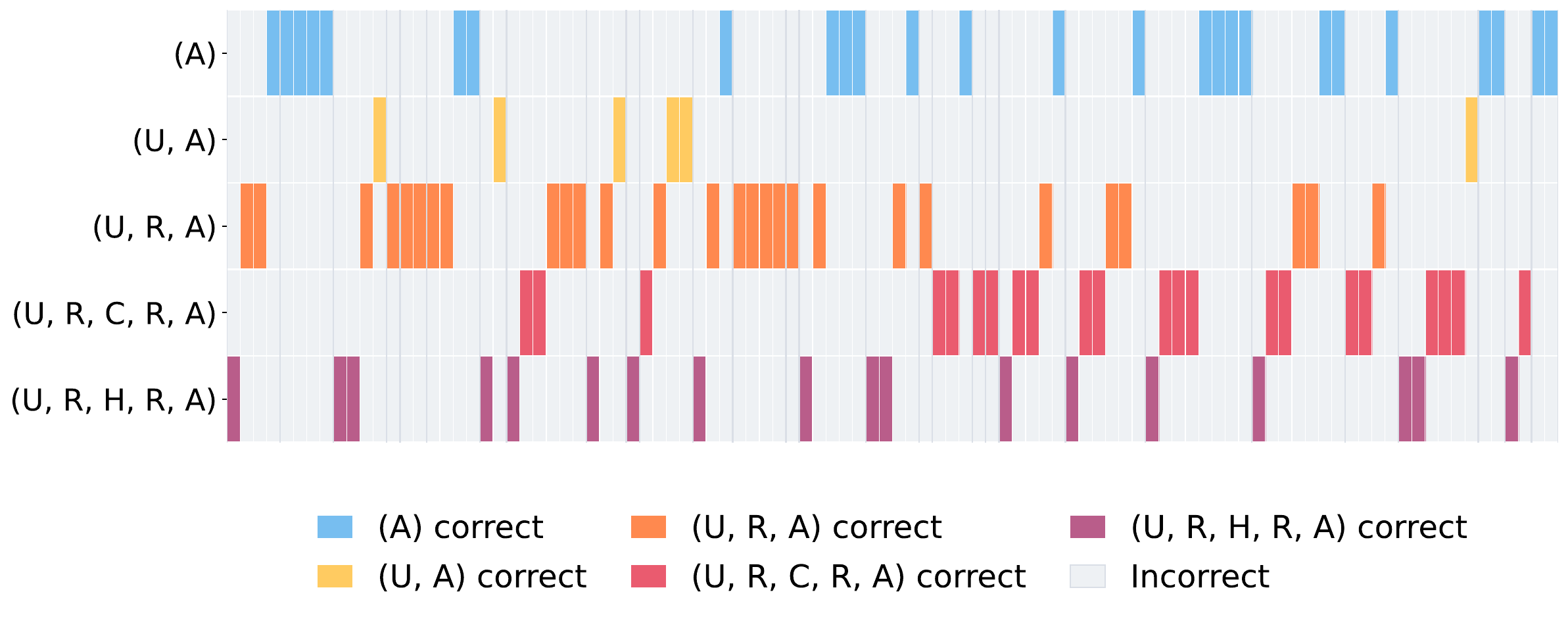}
      \caption{Instance-level path correctness on MMMU.}
      \label{fig:intro_question_correctness}
    \end{subfigure}
  \end{minipage}
  \caption{\textbf{Coordination-path diversity in unified multimodal models.}
  Different coordination paths exhibit complementary strengths across inputs. The large oracle gap over fixed strategies suggests that exploiting coordination-path diversity can significantly improve UMM performance.}
  \label{fig:intro_motivation}
  \vspace{-.1in}
\end{figure*}

While it is promising to exploit the coordination-path diversity, turning this observation into a practical system raises three challenges.
First, coordination categorization is needed: what kinds of contributions understanding and generation can make, and which forms are appropriate for different inputs?
Without categorization, coordination tends to collapse into a single fixed pattern, ignoring task-specific needs and making costly cooperation less likely to yield matching gains.
Second, even after categorization, we need training data and objectives that enable a single UMM to reliably execute different paths rather than merely follow their surface format. For visual roles, the intermediate state may be an abstract construction or a set of hypotheses, so supervision should not require every such step to become a complete image. Third, learning a path planner is a generalization problem under scarce supervision. Labels are expensive to obtain, domain biases vary significantly across datasets, and even with dataset-level knowledge, accurate instance-level path selection remains difficult.

In this paper, we propose \textbf{\method}, a planner-executor framework for adaptive coordination. We first abstract recurring operations in prior multimodal reasoning systems~\citep{goyal2017making,lu2022learn,qin2025uni,cheng2025visualthoughts,cheng2025comt,zhang2025adaptive,adloop,irg,fang2025got} into five functional roles: understanding, reasoning, construction, hypothesis, and answer. To keep the search space trainable, we define five representative coordination paths, each centered on one core role: answering directly, adding explicit understanding, adding textual reasoning, constructing a visual thought, or exploring hypotheses. We then train a path-conditioned executor on role-aligned trajectories so the same UMM can follow different reasoning paths. For visual roles, we use aligned visual thought: the trace remains readable text, while the hidden states of visual-thought spans are supervised by visual summaries. Finally, a planner selects an input-dependent path, and a lightweight query-form calibration step combines learned path scores with simple structural priors.

Our contributions are threefold. (1) We formulate UMM reasoning as coordination-path selection and empirically show strong path diversity across subjects and instances.
(2) We introduce a compact role/path space and train a path-conditioned executor with aligned visual thought, enabling one UMM to realize multiple coordination behaviors.
(3) We build a planner-executor system that selects paths per input, improving accuracy with lower token cost while producing interpretable reasoning traces.

\section{Related Work}
\textbf{Unified Multimodal Models.}
UMMs aim to integrate understanding and generation within a single architecture~\citep{yin2024survey, zhao2025unified}. Recent advances span a diverse set of designs, ranging from models that treat multimodal inputs as unified token sequences for next-token prediction~\citep{wang2024emu3, team2023gemini, bai2025qwen3}, to approaches that incorporate diffusion or flow-based components for improved visual synthesis~\citep{xie2024show, xie2025show, wu2024janus, chen2025janus, ma2024janusflow}, as well as systems that explore different design choices to balance efficiency, scalability, and generation quality~\citep{yang2025mmada, wang2026deepgen, wu2025openuni}. 
Despite their differences, these models share a common goal of capability unification, i.e., equipping a single model with multiple multimodal functionalities. However, multimodal reasoning is largely handled implicitly within the model, without explicit mechanisms to coordinate understanding and generation during inference. This often leads to inconsistencies between the two capabilities~\citep{torchumm}, revealing a gap between unified capabilities and structured reasoning.
Recent work argues that structural unification alone does not ensure functional synergy~\citep{luo2026beyond}. FlexUMM~\citep{luo2026flexumm} addresses this issue through adaptive allocation of shared and specialized computation. Our work studies a complementary level of adaptation: selecting the functional roles and reasoning path used to solve each input.

\textbf{Coordinating Understanding and Generation.}
Role-based coordination and self-play have received increasing attention in multimodal learning~\citep{su2025unigame,he2026visplay}.
Some work couples the two processes during training such as self-play frameworks~\mbox{\citep{su2025unigame}} and reconstruction alignment~\mbox{\citep{xie2025reconstruction}}. They improve global consistency between perception and synthesis, but do not specify how the two capabilities should be coordinated at inference time.
LatentUMM~\citep{luo2026latentumm} aligns cross-modal representations and transformations used for generation and re-encoding to improve consistency. These training approaches improve representations or learning signals, while our framework selects among coordination paths at inference time.
Another direction extends chain-of-thought reasoning to multimodal settings, where intermediate visual representations may influence reasoning~\citep{qin2025uni}. However, the coordination structure is still largely predetermined by the prompting or training format, making it difficult to adapt the amount and type of coordination to each input.
More recent methods introduce explicit coordination mechanisms, such as iterative analyzing-drafting loops~\citep{adloop} and interleaving reasoning and generation for iterative refinement~\citep{irg}. While they integrate generation into the reasoning process during inference, they rely on fixed coordination patterns and do not explicitly distinguish which functional roles are needed for different inputs.
In contrast, we model understanding-generation coordination as path-based coordination: the system first selects a coordination path, then executes the corresponding role sequence. This shifts the focus from designing a single universal coordination protocol to adaptive exploitation.

\section{Methodology}
\label{sec:method}

\begin{figure*}[t!]
  \centering
  \includegraphics[width=1\linewidth]{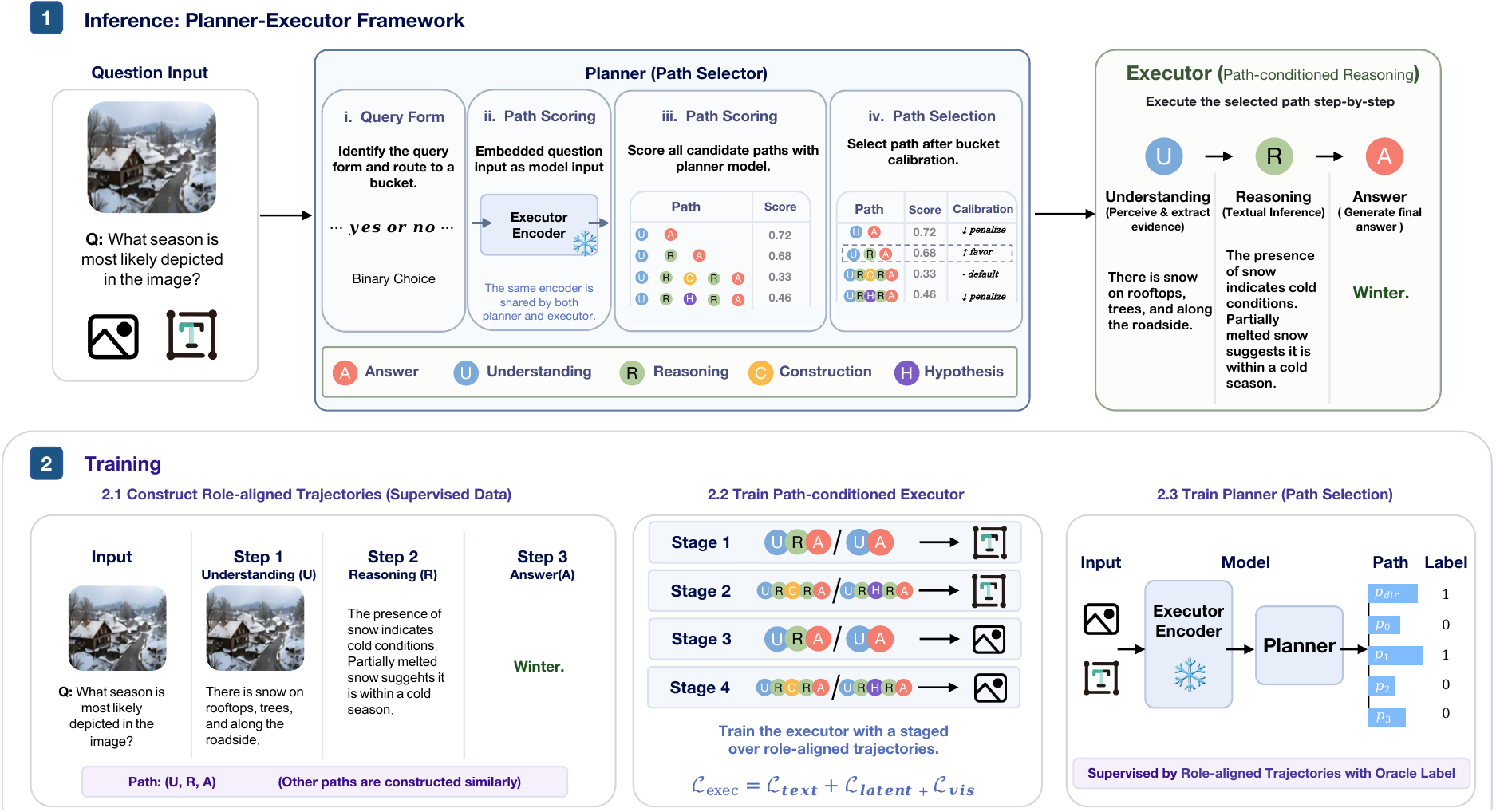}
  \caption{Overview of the training and inference process of \method.}
  \label{fig:overview}
\end{figure*}

\subsection{Problem Formulation} 
\label{sec:problem}

We denote the input of a UMM as $x=(q,\mathcal{I})$, where $q$ is the textual question or instruction and $\mathcal{I}$ is the input image. Given $x$, the model can perform perceptual understanding and generative operations. While both capabilities are available, different inputs may benefit from different ways to organize them, raising a key challenge: how to represent multiple coordination patterns and select an appropriate one for each input?
We address this by formulating understanding-generation coordination as path-based coordination. Instead of directly mapping $x$ to an output $y$, we introduce a coordination path $p$ that specifies a structured coordination strategy. Executing $p$ produces intermediate states that lead to the final output $y$. This formulation avoids assuming a fixed coordination pattern and instead provides a unified interface for representing different strategies within a single model.
Formally, for a path space $\mathcal{P}$, the planner is a path selector $\mathcal{G}_\psi$ that returns a single path before execution:
\begin{equation}
\hat{p}=\mathcal{G}_\psi(x)\in\mathcal{P}, \qquad
y = \mathcal{E}_\theta(x, \hat{p}),
\end{equation}
where $\hat{p}$ denotes the selected path and $\mathcal{E}_\theta$ is the executor that follows this path.
The executor is the UMM itself after path-conditioned training: it receives both the original input and the selected path, then generates the corresponding trace and final output. The planner is a lightweight routing module that selects the path before the UMM executes it.


\subsection{Coordination Categorization}

We model understanding-generation coordination through a set of structured coordination paths. The key idea is to make coordination explicit at the level of \emph{what role each step plays}, rather than treating a trajectory as an arbitrary sequence of tokens. This lets us compare, train, and select different ways of using understanding and generation during inference.

\textbf{Functional roles.}
Different inputs require different uses of the capabilities. For example, one input mainly needs visual evidence understanding, while another may need comparison among possible visual hypotheses. We therefore categorize coordination by the functional role.
The role design is motivated by recurring patterns in existing multimodal reasoning systems. Visual question answering emphasizes explicit understanding~\citep{goyal2017making}. Multimodal chain-of-thought separates textual reasoning~\citep{lu2022learn,qin2025uni}. Interleaved understanding-generation methods suggest that generation can serve roles such as intermediate construction or hypothesis exploration~\citep{adloop,irg,fang2025got}.
Abstracting these observations, we use five functional roles: understanding (U), reasoning (R), construction (C), hypothesis (H), and answer (A). U extracts observations from the input, R performs textual reasoning, and A produces the final answer. C and H are visual-thought roles: construction creates a visual thought for the next step, while hypothesis maintains candidate visual thoughts for comparison. This role set is not intended to be exhaustive. Instead, it provides a compact interface that captures common useful functions while remaining simple enough to train and support path selection.

\textbf{Coordination path space.}
Given these roles, coordination can be viewed as selecting among different coordination paths. Enumerating every role sequence would create a large search space with weak supervision and many redundant variants. We instead define a compact set of representative paths. Each path is centered on one core role, with only the surrounding steps needed to make the path executable. This keeps the space small enough to train and evaluate while still covering qualitatively different coordination patterns: $
\mathcal{P}=\{p_{\mathrm{A}},p_{\mathrm{U}},p_{\mathrm{R}},p_{\mathrm{C}},p_{\mathrm{H}}\},
$
where
\begin{equation*}
\underbrace{p_{\mathrm{A}} = (\mathbf{A})}_{\scriptsize \text{Direct \textbf{A}nswering}},~
\underbrace{p_{\mathrm{U}} = (\mathbf{U}, A)}_{\scriptsize \text{Explicit \textbf{U}nderstanding}},~
\underbrace{p_{\mathrm{R}} = (U, \mathbf{R}, A)}_{\scriptsize \text{Textual \textbf{R}easoning}},~
\underbrace{p_{\mathrm{C}} = (U, R, \mathbf{C}, R, A)}_{\scriptsize \text{Visual-thought \textbf{C}onstruction}},~
\underbrace{p_{\mathrm{H}} = (U, R, \mathbf{H}, R, A)}_{\scriptsize \text{\textbf{H}ypothesis exploration}}.
\end{equation*}

\subsection{Planner-Executor Framework}
\label{sec:planner_executor}

We instantiate path-based coordination with a planner-executor framework. The planner implements $\mathcal{G}_\psi$, selecting a coordination path $\hat{p} \in \mathcal{P}$ conditioned on the input $x$. The executor $\mathcal{E}_\theta$ is the UMM that follows the selected path and returns the intermediate states and final output $y$. 

\textbf{Role-aligned trajectories.}
Training the executor aims to follow a selected path and make intermediate states useful for the next step. We therefore convert heterogeneous examples into \textbf{role-aligned trajectories}. Each trajectory contains the input $x$, a path label $p$, and segments arranged in the role order specified by $p$. 
We use tagged text to mark each role in the trace (e.g., \texttt{Understanding} for U). 
For paths with visual-thought roles, the tagged \texttt{Visual}/\texttt{Hypothesis} span remains readable text, while its hidden states are aligned to a visual summary. 
This provides a lightweight coordination channel that passes visual information to subsequent reasoning, while avoiding the high cost and inaccuracy of explicit image generation and the granularity mismatch introduced by raw visual latent insertion.
Additionally, further analysis of aligned visual thoughts is provided in Appendix~\ref{app:aligned_visual_thought_analysis}, the prompt-level wrappers used at evaluation time are list in Appendix~\ref{app:path_prompts}, and representative trajectories are provided in Appendix~\ref{app:trajectory_examples}.


\textbf{Executor training.}
Given a selected path, the executor must follow the role-tagged interface and make each intermediate state meaningful. 
A single mixed objective can make this difficult because path following, answer prediction, final image generation, and visual-thought alignment impose different signals.
We therefore train the executor with a staged curriculum over role-aligned trajectories. The final run follows a four-stage LoRA chain: textual understanding, visual-thought understanding, plain image answering, and image answering with visual-thought supervision. 
Implementation details are in Appendix~\ref{app:executor_training_details}.
Specifically, each trajectory provides an input $x$, a path label $p$, and target text tokens $z=(z_1,\ldots,z_T)$ for the textual roles. We optimize a role-weighted language modeling loss:
\begin{equation}
\mathcal{L}_{\mathrm{text}}
= - \frac{1}{\sum_{t=1}^{T} w_t}
\sum_{t=1}^{T} w_t \log \pi_\theta(z_t \mid x,p,z_{<t}).
\end{equation}
Here, $\pi_\theta$ is the executor's token distribution and $w_t$ are role-dependent token weights, allowing the same sequence interface to supervise understanding, reasoning, and answer tokens without requiring a separate objective for each role.
For paths with construction or hypothesis roles, each \texttt{Visual}/\texttt{Hypothesis} segment is trained as an aligned visual thought. Let $\bar{h}_j$ denote pooled hidden representation over the $j$-th visual-thought span, and let $v_j$ denote the visual summary embedded from the corresponding reference image. A lightweight projection head $g_\phi$ aligns the executor state to this target:
\begin{equation}
\mathcal{L}_{\mathrm{vis}}
= \frac{1}{J}\sum_{j=1}^{J}\left\|g_\phi(\bar{h}_j)-v_j\right\|_2^2 .
\end{equation}
For trajectories whose final answer is an image, we also keep BAGEL's final image-latent reconstruction loss $\mathcal{L}_{\mathrm{latent}}$. The objective for the executor is
\begin{equation}
\mathcal{L}_{\mathrm{exec}}
= \lambda_{\mathrm{text}}\mathcal{L}_{\mathrm{text}}
+ \lambda_{\mathrm{mse}}\mathcal{L}_{\mathrm{latent}}
+ \lambda_{\mathrm{vis}}\mathcal{L}_{\mathrm{vis}}.
\end{equation}
Terms that are not present in a trajectory, such as visual-thought supervision for $p_{\mathrm{U}}/p_{\mathrm{R}}$ or final image reconstruction for answer-only examples, are omitted. The $\lambda$ coefficients balance text, final-image, and visual-thought supervision.

\textbf{Planner supervision.}
The planner is trained to predict which paths lead to correct outcomes (detail in Sec.~\ref{sec:experimental_setup}). For each input, this yields binary outcomes $r_p\in\{0,1\}$ for paths $p\in\mathcal{P}$. The learned planner produces a path-wise score
\begin{equation}
a_\psi(x,p)=\sigma(f_\psi(x)_p),
\end{equation}
which estimates the probability that path $p$ will succeed on input $x$, with $\sigma$ denoting the sigmoid function.
Since multiple paths can solve the same input, we train the planner as a multi-label predictor rather than imposing a single best-path target. For a minibatch $\mathcal{B}$, the objective is a weighted binary cross-entropy with regularization:
\begin{equation}
\mathcal{L}_{\mathrm{plan}}
=
\frac{1}{\sum_{i\in\mathcal{B}}\omega_i}
\sum_{i\in\mathcal{B}}\omega_i
\frac{1}{|\mathcal{P}|}
\sum_{p\in\mathcal{P}}
\beta_{i,p}\,
\mathrm{BCEWithLogits}\!\left(f_\psi(x_i)_p, r_{i,p}\right)
+ \lambda_{\mathrm{reg}}\mathcal{R}(\psi).
\end{equation}
Here, $\omega_i$ is a sample weight and $\beta_{i,p}$ is a path-level label weight. In practice, samples with fewer successful paths receive larger weight because they provide sharper routing supervision, and positive labels outside $p_{\mathrm{A}}$ are mildly upweighted to reduce collapse to $p_{\mathrm{A}}$. $\mathcal{R}(\psi)$ denotes standard planner regularization, implemented as weight decay on the planner parameters. This objective preserves the multi-path nature of the supervision. The final single path is chosen only after query-form calibration.

\textbf{Query-form calibrated path selection.}
At inference time, directly selecting the path with the highest predicted score can be unstable, as the planner must generalize across dataset-specific domain biases under limited supervision. We therefore add a query-form prior based on surface structure. This prior does not replace the planner. Instead, it calibrates planner scores using cues that often correlate with the required coordination. For example, simple counting or binary-choice questions tend to favor simple paths, while geometry or chart reasoning may benefit from more structured coordination.
Concretely, we introduce a lightweight calibration mechanism that adjusts path selection based on query form. Inputs are grouped into coarse query-form buckets using simple surface patterns rather than dataset identity. For each bucket, we apply temperature scaling and path-specific biases to the planner scores, and select a path only when its advantage over a default path exceeds a margin.

\section{Experiments}


\subsection{Experimental Setup}
\label{sec:experimental_setup}

\textbf{Backbone and Training.}
We instantiate the executor with BAGEL~\citep{deng2025emerging} and train lightweight LoRA adapters~\citep{hu2022lora, peft} for path execution. Evaluation is conducted with TorchUMM~\citep{torchumm} for fair comparison. 
For executor training, we train BAGEL on path-aligned trajectories with supervision across the coordination paths in Sec.~\ref{sec:problem}.
Notably, our executor uses a comparatively \textit{smaller} training set that shows our empirical gains come from exploiting the right form of understanding-generation coordination, not simply from using a larger post-training corpus.
Executor training is organized into four staged splits that activate different links of the path.
We report answer accuracy, format accuracy, CE, visual-thought alignment loss, and image-latent MSE where applicable, with staged training diagnostics provided in Appendix~\ref{app:executor_diagnostics}.
For planner training, the supervision is built after executor training by running all five candidate paths on roughly 8k calibration examples and recording which paths solve each query. 
More training details and results are in Appendix~\ref{app:training_data_details}, \ref{app:executor_training_details}, \ref{app:planner_configuration}, and ~\ref{app:more_experimental_results}.

\textbf{Planner calibration.}
We treat bucket construction as a calibration step rather than a fully hand-written procedure. 
Buckets and routing rules are derived from an auxiliary calibration pool, including the planner-construction split, the MMBench validation split, a subset of 
MathVerse~\citep{zhang2024mathverse}.
GPT-5.5 is used for summarization, identifying recurring query-form patterns, which we consolidate into shared query-form buckets and path-based rules. 
At evaluation time, the planner relies only on the input query form and learned path scores.

\textbf{Benchmarks.}
We evaluate understanding on MMMU~\citep{yue2024mmmu}, MMBench-EN/CN~\citep{liu2024mmbench}, MathVista~\citep{lu2023mathvista}, and MMStar~\citep{chen2024mmstar}, covering expert knowledge, cross-lingual visual QA, mathematical reasoning, and fine-grained visual reasoning. We evaluate generation on GenEval~\citep{ghosh2023geneval} and WISE~\citep{niu2025wise}, and evaluate understanding-generation consistency on UnifiedBench~\citep{yan2025can}. For generation and consistency benchmarks, the trained executor is run under the original query input, so these results show the effects of executor training and aligned visual-thought supervision rather than input-dependent routing. Detailed benchmark descriptions and scoring protocols are provided in Appendix~\ref{app:benchmark_metric_details}.

\textbf{Baselines.}
We compare against unified multimodal generation/understanding models when reported under the same benchmark protocol, including BAGEL~\citep{deng2025emerging}, BLIP3-o~\citep{chen2025blip3}, Janus/Janus-Pro~\citep{wu2024janus, chen2025janus}, JanusFlow~\citep{ma2024janusflow}, Show-o/Show-o2~\citep{xie2024show, xie2025show}, Emu3~\citep{wang2024emu3}, Emu3.5~\citep{cui2025emu3}, OmniGen2~\citep{wu2025omnigen2}, TokenFlow~\citep{geyer2023tokenflow}, DeepGen~\citep{wang2026deepgen}, MMaDA~\citep{yang2025mmada}, and Ovis-U1~\citep{wang2025ovisu1}. We also compare with BAGEL-based or unified-reasoning variants including RecA~\citep{xie2025reconstruction}, UniGame~\citep{su2025unigame}, IRG~\citep{irg}, UniCoT~\citep{qin2025uni} and AD-Loop~\citep{adloop}. 

\subsection{Main Results}

\begin{table*}[t]
\centering
\scriptsize
\setlength{\tabcolsep}{2.7pt}
\renewcommand{\arraystretch}{1.04}
\caption{Main results on understanding and generation benchmarks. Missing entries indicate unavailable results. Relative-gain rows report improvements over the corresponding BAGEL baseline, with AD-Loop computed from the values reported in its paper.}
\vspace{-.1in}
\label{tab:understanding_main}
\resizebox{.85\textwidth}{!}{
\begin{tabular}{llcccccc!{\vrule width 0.9pt}cc}
\toprule
\multicolumn{2}{c}{} & \multicolumn{6}{c}{Understanding} & \multicolumn{2}{c}{Generation} \\
\cmidrule(lr){3-8}\cmidrule(lr){9-10}
Method & Params & MMMU & MMB-EN & MMB-CN & MathVista & MMStar & Avg. & GenEval & WISE \\
\midrule
BAGEL & 7B+7B & 51.90 & 82.65 & 80.94 & 71.60 & 63.20 & 70.06 & 78.81 & 0.3989 \\
Janus-Pro & 7B & 40.70 & 67.64 & 64.94 & 42.80 & 42.00 & 51.62 & 78.92 & 0.3811 \\
Janus & 1.3B & 27.30 & 27.69 & 37.19 & 26.60 & 26.67 & 29.09 & 40.04 & 0.2222 \\
JanusFlow & 1.3B & 29.00 & 64.16 & 60.31 & 34.80 & 39.13 & 45.48 & 49.99 & 0.2954 \\
Show-o2 & 7B & 47.90 & 77.25 & 76.68 & 51.50 & 54.53 & 61.57 & 59.87 & 0.3595 \\
Show-o2 & 1.5B & 37.10 & 65.09 & 61.04 & 37.90 & 43.87 & 49.00 & 55.49 & 0.3349 \\
Show-o & 1.3B & 26.10 & 43.48 & 11.01 & 29.00 & 27.87 & 27.49 & 65.06 & 0.3037 \\
Emu3 & 8B & 31.40 & 59.53 & 45.66 & 44.90 & 43.60 & 45.02 & 45.76 & 0.3373 \\
OmniGen2 & 3B+4B & 46.00 & 76.36 & 75.43 & 63.50 & 52.47 & 62.75 & 78.53 & 0.4029 \\
MMaDA & 7B & 28.90 & 28.16 & 20.57 & 24.90 & 31.73 & 26.85 & 46.12 & \textbf{0.6560} \\
Ovis-U1 & 3B & 43.70 & 78.55 & 76.99 & 68.50 & 58.27 & 65.20 & \textbf{90.05} & 0.3755 \\
Emu3.5 & 34B & 29.20 & 17.56 & 18.13 & 41.67 & 30.27 & 27.37 & 81.83 & \underline{0.6331} \\
\midrule
BLIP3-o & 4B & -- & -- & -- & -- & -- & -- & 81.36 & 0.4138 \\
TokenFlow & 7B+14B & -- & -- & -- & -- & -- & -- & 52.21 & 0.3056 \\
DeepGen & 3B+2B & -- & -- & -- & -- & -- & -- & \underline{86.59} & 0.5470 \\
\midrule
RecA & -- & 52.30 & 82.65 & 80.94 & 51.60 & 68.47 & 67.19 & 83.05 & 0.4225 \\
UniGame & -- & 52.40 & 82.75 & 80.94 & \underline{72.20} & \underline{68.73} & 71.40 & 86.17 & 0.4032 \\
IRG & -- & 48.00 & 60.47 & 57.35 & 68.00 & 61.53 & 59.07 & 72.06 & 0.3842 \\
UniCoT & -- & \underline{53.10} & \underline{83.12} & \underline{80.99} & \textbf{73.00} & \textbf{70.00} & \underline{72.04} & 77.91 & 0.4037 \\
\rowcolor{oursrow}
Ours & -- & \textbf{54.11} & \textbf{86.31} & \textbf{83.57} & \underline{72.20} & 68.07 & \textbf{72.85} & 80.00 & 0.4100 \\
\midrule
\multicolumn{2}{l}{AD-Loop rel. gain} & +3.6 & +3.1 & -- & \textbf{+4.9} & +5.5 & -- & -- & -- \\
\rowcolor{oursrow}
\multicolumn{2}{l}{Ours rel. gain} & \textbf{+4.3} & \textbf{+4.4} & \textbf{+3.2} & +0.8 & \textbf{+7.7} & -- & -- & -- \\
\bottomrule
\end{tabular}}
\vspace{-.1in}
\end{table*}

\Cref{tab:understanding_main} summarizes the main comparison across understanding and generation. On understanding benchmarks, our method consistently improves over the BAGEL backbone across all datasets. For example, it achieves gains of +4.3\% on MMMU and +4.4\% on MMBench-EN, and shows a particularly strong improvement of +7.7\% on MMStar.
The improvement on MathVista is comparatively smaller (+0.8\%), which we attribute to the relatively homogeneous problem types in this benchmark. In such cases, path diversity provides limited additional benefit, and the similarity of inputs also makes it more challenging for the planner to reliably distinguish between suitable coordination paths.
In contrast, datasets such as MMMU and MMStar contain more diverse inputs, where different coordination strategies can be more effectively exploited.
Compared with BAGEL-based and unified-reasoning variants, our method achieves the best results on MMMU, MMBench-EN, and MMBench-CN, while remaining competitive on MathVista and MMStar. This supports our central claim that exploiting coordination-path diversity is more effective than enforcing a single fixed coordination pattern.

The two rightmost columns report generation benchmarks. 
Since generation uses the trained executor without planner routing, these results reflect executor training rather than adaptive path selection.
Ours improves the BAGEL backbone from 78.81 to 80.00 on GenEval (+1.5\% relative gain) and from 0.3989 to 0.4100 on WISE (+2.8\% relative gain).
Compared with other post-training methods, our approach achieves competitive or better performance on these benchmarks.
The gains indicate that aligned visual-thought supervision can improve or preserve generation ability while supporting the understanding-side coordination paths. 
Complete category-level generation results of each benchmark are provided in Appendix~\ref{app:generation_full}.

\subsection{Understanding-Generation Consistency}

\begin{wraptable}{r}{0.48\textwidth}
\vspace{-0.1in}
\centering
\caption{Understanding-generation consistency on UnifiedBench. Higher similarities are better.}
\vspace{-.1in}
\label{tab:unifiedbench}
\resizebox{0.48\textwidth}{!}{
\begin{tabular}{lccccc}
\toprule
Method & CLIP & DINOv2 & DINOv3 & LongCLIP & Overall \\
\midrule
BAGEL & 0.8947 & 0.7877 & 0.7240 & 0.9321 & 0.8346 \\
Ours & 0.8958 & 0.7865 & 0.7338 & 0.9358 & 0.8380 \\
Ours rel. gain & +0.1\% & -0.2\% & +1.4\% & +0.4\% & +0.4\% \\
\bottomrule
\end{tabular}}
\vspace{-0.1in}
\end{wraptable}

To evaluate understanding-generation consistency, we use UnifiedBench~\citep{yan2025can}. This diagnostic complements the standalone understanding and generation benchmarks by testing whether executor training preserves information across a reconstruction loop. We compare against the BAGEL backbone and report both absolute scores and relative gains.
Table~\ref{tab:unifiedbench} shows that our executor improves the BAGEL backbone from 0.8346 to 0.8380 overall. 
This positive improvement supports the intended effect of aligned visual-thought supervision, where image-derived supervision encourages textual visual thoughts to carry visual information that remains useful when the model later regenerates an image.
Appendix~\ref{app:aligned_visual_thought_analysis} further analyzes this design by comparing it with explicit visual latent feedback and complete image feedback.


\subsection{Planner Analysis}
\begin{figure*}[t]
  \centering
  \includegraphics[width=\textwidth]{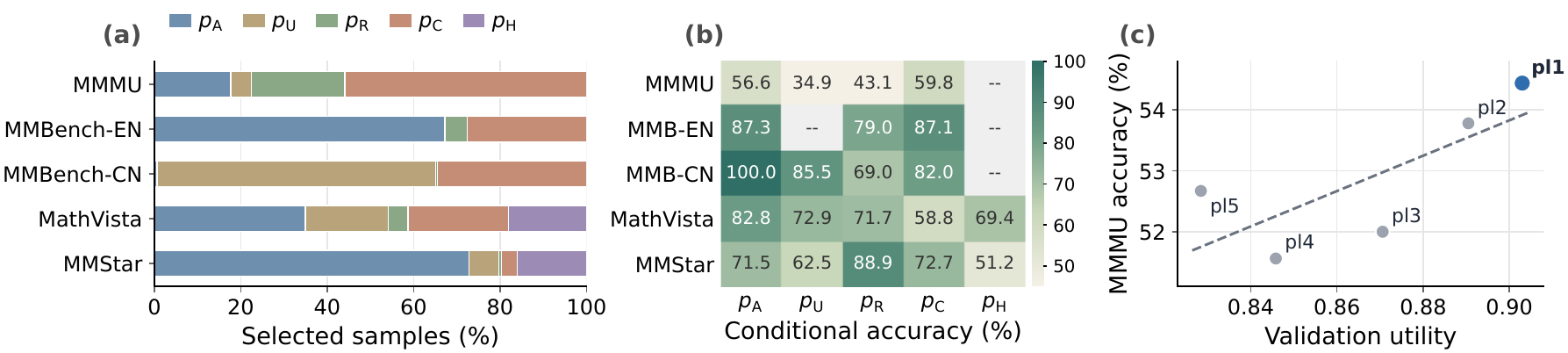}
  \caption{Planner behavior across benchmarks and validation transfer. (a) Selected path distribution. (b) Conditional accuracy on samples selected for each path. Missing entries indicate paths selected for zero samples. (c) Planner training validation utility versus routed MMMU accuracy for five planner checkpoints or configurations.}
  \label{fig:planner_dataset_behavior}
\end{figure*}

\textbf{Planner behavior across benchmarks.}
Figure~\ref{fig:planner_dataset_behavior} reports three views of the planner. Panel (a) shows the fraction of questions assigned to each path across five understanding benchmarks. The selected path distribution varies across datasets, suggesting that the planner is not simply applying a fixed preference for deeper reasoning. The planner assigns most MMMU examples to $p_{\mathrm{C}}$, consistent with its expert-level questions where intermediate visual-thought construction and structured reasoning are often useful. In contrast, MMBench-EN and MMStar are dominated by $p_{\mathrm{A}}$, reflecting their larger fraction of recognition, commonsense, and option-matching questions where direct answering is usually sufficient. MathVista shows a more balanced path distribution. This may reflect the relatively homogeneous problem types in the benchmark, where coordination patterns are less distinct and harder to separate.
Panel (b) reports conditional accuracy within the examples selected for each path, showing that non-dominant paths can still achieve reasonable accuracy on the subsets. This indicates that the planner does not collapse to a single dominant strategy, but instead distributes inputs across multiple paths. At the same time, the comparable accuracy across paths suggests that the path design is broadly well-aligned with different types of inputs.
Panel (c) compares planner training validation utility against routed MMMU accuracy for five planner versions. Higher validation utility generally corresponds to better downstream routing, and the final planner achieves both the highest validation utility and the best MMMU accuracy. 

\textbf{Planner ablations.}
Table~\ref{tab:planner_ablation} ablates fixed paths, random selection, Model, Bucket, BAGEL path choice, our planner, and two reference settings. Model uses only the learned planner scores, while Bucket uses only query-form bucket rules to select a fixed path without learned scores.
These subset results are intended to diagnose planner behavior rather than replace the full benchmark results in Table~\ref{tab:understanding_main}.
The ablation highlights three points. First, model scores and query-form buckets are both useful, but neither is sufficient alone: raw scores can collapse toward a narrow preference, while bucket-only routing misses fine-grained distinctions. Although Model or Bucket can be competitive on individual datasets, our planner is better or tied on four of five benchmarks against each variant and obtains the best average among deployable planners. Second, dataset-adapted calibration can be stronger when the target benchmark distribution is known, but this sacrifices generalizability. Our planner gives up some target-specific headroom in exchange for a shared query-form calibration policy. Third, BAGEL's own path choice remains below our planner, indicating that path selection is not automatically solved by a capable UMM and requires explicit planner design.

\begin{table*}[t!]
\centering
\scriptsize
\setlength{\tabcolsep}{3.2pt}
\renewcommand{\arraystretch}{1.08}
\caption{Planner analysis and ablations on understanding benchmarks. For this ablation, MMMU is evaluated on the full set, while MMBench-EN, MMBench-CN, MathVista, and MMStar use 200-example subsets to keep the cost of evaluating a wide range of planner variants manageable. }
\label{tab:planner_ablation}
\resizebox{\textwidth}{!}{
\begin{tabular}{lcccccccccc|cc}
\toprule
Dataset & $p_{\mathrm{A}}$ & $p_{\mathrm{U}}$ & $p_{\mathrm{R}}$ & $p_{\mathrm{C}}$ & $p_{\mathrm{H}}$ & Rand. & Model & Bucket & BAGEL & Ours & Adapted & Oracle \\
\midrule
MMMU & 51.89 & 50.11 & 52.22 & 52.33 & 51.00 & 49.78 & 52.22 & 51.78 & 52.33 & \cellcolor{oursrow}54.11 & 56.11 & 72.00 \\
MMBench-EN & 92.00 & 90.00 & 90.00 & 90.00 & 89.00 & 90.00 & 90.00 & 92.00 & 89.00 & \cellcolor{oursrow}92.00 & 94.50 & 98.00 \\
MMBench-CN & 88.00 & 86.50 & 85.00 & 83.00 & 86.00 & 82.00 & 83.00 & 86.50 & 87.00 & \cellcolor{oursrow}85.00 & 91.00 & 97.00 \\
MathVista & 67.50 & 65.50 & 67.50 & 72.00 & 67.50 & 66.50 & 69.00 & 66.50 & 67.00 & \cellcolor{oursrow}68.50 & 77.50 & 88.00 \\
MMStar & 65.50 & 60.00 & 62.50 & 54.50 & 58.00 & 62.00 & 58.00 & 69.50 & 62.00 & \cellcolor{oursrow}70.00 & 74.50 & 84.00 \\
\midrule
Average & 72.98 & 70.42 & 71.44 & 70.37 & 70.30 & 70.06 & 70.44 & 73.26 & 71.47 & \cellcolor{oursrow}73.99 & 78.72 & 87.80 \\
\bottomrule
\end{tabular}}
\end{table*}

Additionally, we analyze the planner from several perspectives and provide more discussion in the appendix. Appendix~\ref{app:planner_training_transfer} shows the path distributions of planner checkpoints, and Appendix~\ref{app:planner_feature_space} visualizes the planner feature space, showing why domain shift and overlapping path labels make planner generalization difficult.

\subsection{Token-Accuracy Tradeoff}
\begin{figure*}[t]
  \centering
  \includegraphics[width=\textwidth]{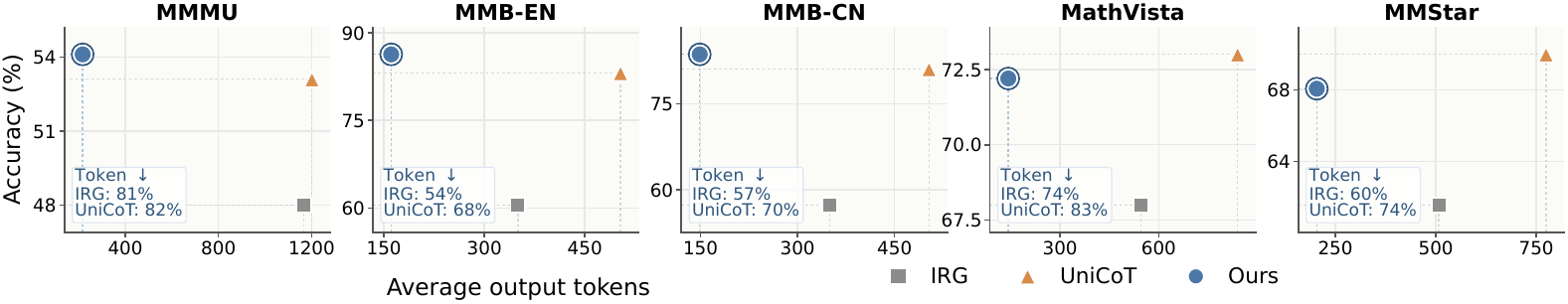}
  \caption{Accuracy versus average output-token cost on understanding benchmarks.}
  \label{fig:token_accuracy_tradeoff}
\end{figure*}

Adaptive coordination improves accuracy without simply increasing the amount of generated reasoning. Figure~\ref{fig:token_accuracy_tradeoff} compares average output-token cost and accuracy against post-training reasoning methods. Across MMMU, MMBench-EN, MMBench-CN, MathVista, and MMStar, our method is consistently closer to the upper-left region. It uses substantially fewer output tokens while matching or improving accuracy on most benchmarks. This supports that the gain comes from invoking the appropriate path and roles when useful, rather than forcing every query through a long reasoning trace. The blue annotations report the token reduction of our method relative to IRG and UniCoT. We omit AD-Loop from this cost comparison because the per-example outputs required for token accounting are unavailable.

\newtcolorbox{pathcase}[2][]{
  enhanced,
  breakable,
  colback=white,
  colframe=black!13,
  colbacktitle=oursrow,
  coltitle=black,
  title={#2},
  fonttitle=\bfseries\footnotesize,
  fontupper=\scriptsize,
  boxrule=0.45pt,
  arc=1.5pt,
  left=4pt,
  right=4pt,
  top=3pt,
  bottom=3pt,
  before skip=3pt,
  after skip=4pt,
  #1
}
\newcommand{\caseitem}[2]{\textcolor{black!65}{\textsc{#1.}}~#2\par}
\newcommand{\okpred}[1]{\cellcolor{oursrow}\textbf{#1}}
\newcommand{\casepred}[6]{%
\caseitem{Predictions}{%
\begingroup
\setlength{\tabcolsep}{4.0pt}
\renewcommand{\arraystretch}{0.92}
\begin{tabular}[t]{@{}cccccc@{}}
$p_{\mathrm{A}}$ & $p_{\mathrm{U}}$ & $p_{\mathrm{R}}$ & $p_{\mathrm{C}}$ & $p_{\mathrm{H}}$ & GT \\
#1 & #2 & #3 & #4 & #5 & \textbf{#6}
\end{tabular}
\endgroup}}
\newcommand{\casetrace}[1]{\textcolor{black!65}{\textsc{Trace.}}\par #1}
\newcommand{\tracehead}[2]{\textbf{#1:}~#2\par}

\subsection{Qualitative Analysis}
\label{sec:qualitative}

We show example from the full five-path evaluation. Only the highlighted path gives the correct answer, making the role of path-specific execution visible at the question level. Additional examples for the other paths are provided in Appendix~\ref{app:additional_qualitative_examples}.

\begin{pathcase}[borderline west={1.3pt}{0pt}{traj_color}, colframe=traj_color!40!black]{$p_{\mathrm{C}}$: visual-thought construction}
\caseitem{Question}{In a bridge circuit with resistors $1\Omega$, $3\Omega$, $2\Omega$, and $R$, the ammeter reads zero. What is $R$? Options: A. $1.5\Omega$, B. $4\Omega$, C. $5\Omega$, D. $6\Omega$.}
\casepred{A}{C}{C}{\okpred{D}}{B}{D}
\casetrace{\tracehead{Understanding}{A circuit diagram shows a battery connected to four resistors in a diamond shape, with an ammeter in the middle.}
\tracehead{Reasoning}{The circuit is a Wheatstone bridge. For the ammeter to read zero, the resistance ratios must balance, so $1/3=2/R$.}
\tracehead{Construction}{A constructed bridge-state visual thought isolates the diamond bridge, places the ammeter at the center, and labels the four resistor arms.}
\tracehead{Reasoning}{Solving the balanced-arm relation gives $R=6\Omega$.}
\tracehead{Answer}{D}}
\caseitem{Why this path}{The useful intermediate state is a structured visual relation over circuit arms. This is exactly where visual-thought construction helps: it turns a cluttered diagram into the relation needed for subsequent reasoning.}
\end{pathcase}

\section{Conclusion and Limitations}

We study how a unified multimodal model should decide whether, when, and how to coordinate understanding and generation. We introduce a compact path space, train an executor to follow these paths through a unified interface, and use a query-form calibrated planner to select a path for each input.
Our experiments show that different examples favor different paths, and oracle path selection remains far above any fixed path, indicating strong complementarity among coordination patterns. At the same time, raw planner scores, query-form rules alone, and BAGEL's own path choice are insufficient, suggesting that path selection is itself a central modeling problem. Overall, the results support treating coordination policy as a first-class component of unified multimodal reasoning. 

The main limitation is path selection: the deployable planner still leaves a large gap to oracle routing, and learning a planner that generalizes robustly across domains remains challenging. We provide a fuller discussion in Appendix~\ref{app:limitations}.

\begin{ack}
We thank William \& Mary Research Computing for providing computational resources and/or technical support that have contributed to the results reported within this paper. URL: https://www.wm.edu/it/rc.
\end{ack}

\bibliographystyle{plain}
\bibliography{refs}

\appendix

\begin{center}
{\LARGE\bfseries Appendix}
\end{center}

\vspace{1.0em}
\noindent{\Large\bfseries Contents}
\vspace{0.7em}

\newcommand{\appcontentsmain}[3]{%
  \par\noindent
  \hyperref[#1]{\textcolor{blue}{\textbf{#2\hspace{1em}#3}}}%
  \leaders\hbox to .55em{\hfil.\hfil}\hfill
  \textbf{\pageref{#1}}\par
}
\newcommand{\appcontentssub}[3]{%
  \par\noindent\hspace*{2.3em}
  \hyperref[#1]{\textcolor{blue}{#2\hspace{0.8em}#3}}%
  \leaders\hbox to .55em{\hfil.\hfil}\hfill
  \pageref{#1}\par
}

\begingroup
\small
\appcontentsmain{app:path_diversity_full}{A}{Additional Path-Diversity Visualizations}
\appcontentsmain{app:training_data_details}{B}{Training Data Construction}
\appcontentsmain{app:benchmark_metric_details}{C}{Benchmark and Metric Details}
\appcontentsmain{app:aligned_visual_thought_analysis}{D}{Aligned Visual Thought Analysis}
\appcontentsmain{app:planner_details}{E}{More Details and Analysis on Planner}
\appcontentssub{app:planner_configuration}{E.1}{Planner Training Details}
\appcontentssub{app:planner_training_transfer}{E.2}{Planner Training Transfer}
\appcontentssub{app:planner_feature_space}{E.3}{Planner Feature-Space Analysis}
\appcontentsmain{app:more_experimental_results}{F}{More Experimental Results}
\appcontentssub{app:adloop_relative_gain}{F.1}{Relative Gains of Post-training Methods}
\appcontentssub{app:path_format_compliance}{F.2}{Path Format Compliance on MMMU}
\appcontentssub{app:different_backbone}{F.3}{Different Backbone Experiments}
\appcontentssub{app:generation_full}{F.4}{Additional Generation Results}
\appcontentsmain{app:executor_training_details}{G}{Executor Training Details}
\appcontentssub{app:executor_data_stats}{G.1}{Executor Data Statistics}
\appcontentssub{app:executor_diagnostics}{G.2}{Executor Training Diagnostics}
\appcontentsmain{app:path_prompts}{H}{Path Prompt Templates}
\appcontentsmain{app:additional_qualitative_examples}{I}{Additional Qualitative Examples}
\appcontentsmain{app:trajectory_examples}{J}{Role-Aligned Trajectory Examples}
\appcontentsmain{app:limitations}{K}{Limitations}
\appcontentsmain{app:broader_impact}{L}{Broader Impact}
\endgroup

\section{Additional Path-Diversity Visualizations}
\label{app:path_diversity_full}

Figure~\ref{fig:appendix_subject_path_affinity_full} and Figure~\ref{fig:appendix_question_path_correctness_full} provide the full-resolution versions of the compact MMMU visualizations in Figure~\ref{fig:intro_motivation}. The row labels use the role-sequence notation from Sec.~\ref{sec:problem}: $(A)$ corresponds to $p_{\mathrm{A}}$, $(U,A)$ to $p_{\mathrm{U}}$, $(U,R,A)$ to $p_{\mathrm{R}}$, $(U,R,C,R,A)$ to $p_{\mathrm{C}}$, and $(U,R,H,R,A)$ to $p_{\mathrm{H}}$.

\begin{center}
  \centering
  \includegraphics[width=\textwidth]{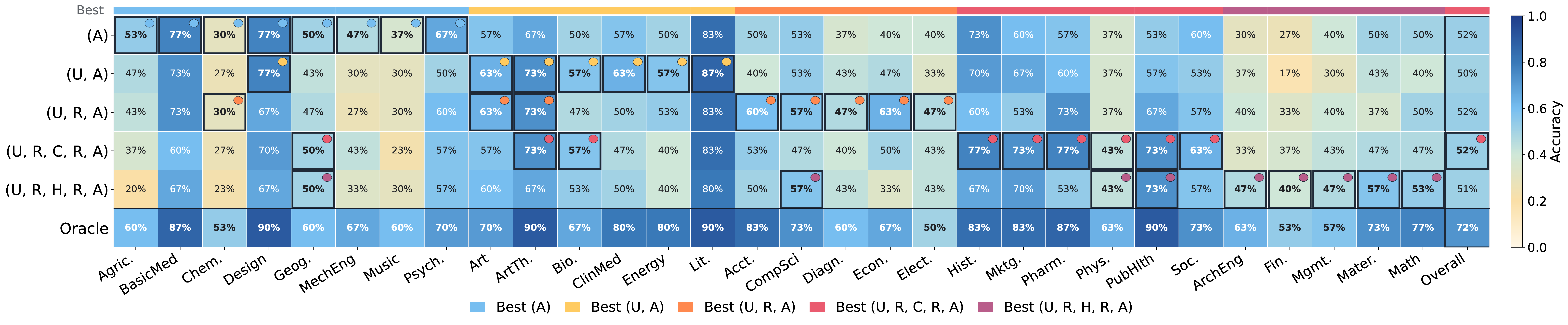}
  \captionsetup{type=figure,hypcap=false}
  \caption{\textbf{Full subject-level path affinity on MMMU.}
  Each column corresponds to an MMMU subject, and each row reports the accuracy of one coordination path. The overall pattern shows that subject domains favor different paths, while the oracle row remains substantially higher than any fixed path, supporting the need for input-dependent path selection.}
  \label{fig:appendix_subject_path_affinity_full}
\end{center}

\begin{center}
  \centering
  \includegraphics[width=\textwidth]{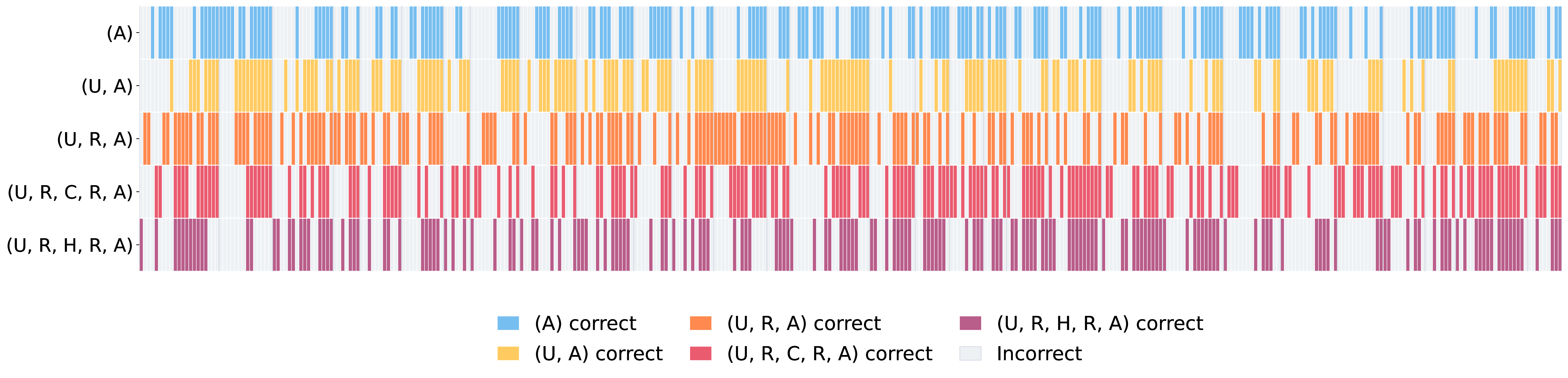}
  \captionsetup{type=figure,hypcap=false}
  \caption{\textbf{Full instance-level path correctness on MMMU.}
  Each column is an MMMU example and each row is a coordination path. Colored cells indicate that the corresponding path answers the example correctly, while gray cells indicate failure. The sparse and non-identical correctness patterns show that path complementarity also appears at the individual-question level, not only after aggregating by subject.}
  \label{fig:appendix_question_path_correctness_full}
\end{center}

\section{Training Data Construction}
\label{app:training_data_details}

We construct path-aligned trajectories from both understanding and generation sources, following the path space in Sec.~\ref{sec:problem}. For $p_{\mathrm{U}}=(U,A)$, we use VQAv2~\citep{goyal2017making} to form perception-only trajectories. For $p_{\mathrm{R}}=(U,R,A)$, we use ScienceQA~\citep{lu2022learn} for understanding and LAION-Aesthetics-High-Resolution-GoT~\citep{fang2025got} for generation. For ScienceQA, we combine questions with answer choices, map the answer index to its option text, and use lecture and solution fields as textual reasoning. For GoT generation data, the prompt defines the generation target and the GoT annotation provides the reasoning trace.

For $p_{\mathrm{C}}=(U,R,C,R,A)$, we construct examples that require intermediate visual-thought construction. On the understanding side, we process CoMT~\citep{cheng2025comt} and CoF-SFT~\citep{zhang2025adaptive}, using provided or zoomed visual references as construction targets. On the generation side, we use FLUX-Reason-6M~\citep{fang2025flux}, where the first reasoning step specifies partial visual elements, a reference image is synthesized, and the second reasoning step completes the remaining elements. For $p_{\mathrm{H}}=(U,R,H,R,A)$, we use GoT generation data and VL-PRM300K~\citep{ong2025training} to create examples with multiple candidate visual hypotheses followed by comparative reasoning.
We sample approximately 5k examples for each nontrivial level and source group. When native intermediate annotations are unavailable, we use GPT-5.4 and Claude Code Opus 4.6 as text teachers to produce role-aligned textual segments, and use BAGEL to synthesize reference images needed for aligned visual-thought supervision. After executor training, planner supervision is constructed by running all five candidate paths on roughly 8k calibration examples and recording which paths solve each query.
As summarized in Table~\ref{tab:training_data_comparison}, our executor uses a comparatively small training set. Our empirical gains therefore come from exploiting the right form of understanding-generation coordination, not simply from using a larger post-training corpus.
\begin{center}
\centering
\scriptsize
\setlength{\tabcolsep}{4.2pt}
\renewcommand{\arraystretch}{1.06}
\captionsetup{type=table,hypcap=false}
\caption{Training data scale and sources.}
\label{tab:training_data_comparison}
\resizebox{0.55\textwidth}{!}{
\begin{tabular}{lrp{0.18\textwidth}}
\toprule
Method & Training data scale & Sources \\
\midrule

RecA & $>$500K & \citep{liu2023visual,zou2026advancing} \\
UniGame & $\sim$450K & \citep{goyal2017making,sharma2018conceptual} \\
UniCoT & $\sim$290K & \citep{ye2025echo,chen2025sharegpt,li2024llava,brejcha2017geopose3k,wasserman2025paint} \\
AD-Loop & $\sim$51K & \citep{cheng2025comt,zhang2025adaptive,shao2024visual,li2025imagine,fang2025got,xiao2025omnigen} \\
IRG & $\sim$300K & \citep{lin2025uniworld,chen2025blip3,jiang2025t2i} \\
Ours & $\sim$38K & \citep{goyal2017making,lu2022learn,fang2025got,cheng2025comt,zhang2025adaptive,fang2025flux,ong2025training} \\
\bottomrule
\end{tabular}}
\end{center}

\section{Benchmark and Metric Details}
\label{app:benchmark_metric_details}

In Table~\ref{tab:understanding_main}, understanding and GenEval scores are accuracies in percent, WISE reports WiScore, and Avg. is the mean over the five understanding benchmarks when all scores are available. Appendix~\ref{app:adloop_relative_gain} reports a separate relative-gain comparison with AD-Loop to avoid mixing potentially different backbone and evaluation settings in the main table.

\textbf{Understanding benchmarks.}
MMMU~\citep{yue2024mmmu} evaluates college-level multimodal knowledge and reasoning. MMBench-EN and MMBench-CN~\citep{liu2024mmbench} evaluate instruction-following visual question answering in English and Chinese, with the Chinese split testing behavior under a language distribution different from our mostly English training sources. MathVista~\citep{lu2023mathvista} emphasizes mathematical and visual reasoning, while MMStar~\citep{chen2024mmstar} focuses on fine-grained multimodal reasoning. We report the official accuracy for each benchmark, and analyze selected subsets and per-path behavior when path-level outputs are available.

\textbf{Generation benchmarks.}
GenEval~\citep{ghosh2023geneval} measures text-to-image object binding and compositional accuracy across categories such as object count, color, position, and attribute binding. WISE~\citep{niu2025wise} evaluates broader text-to-image alignment across domains including culture, time, space, biology, physics, and chemistry. Generation evaluation uses the trained executor under the same generation protocol across prompts, so these scores reflect executor training rather than planner routing.

\textbf{Understanding-generation consistency.}
UnifiedBench~\citep{yan2025can} evaluates whether a UMM preserves image information across a reconstruction loop. The model first converts an input image into text and then regenerates an image from that text. The reconstructed image is compared with the original image using CLIP, DINOv2, DINOv3, and LongCLIP similarities. This benchmark probes the alignment effect of executor training and aligned visual-thought supervision, because success requires visual information encoded during understanding to remain useful for later generation.

\section{Aligned Visual Thought Analysis}
\label{app:aligned_visual_thought_analysis}

We further test whether aligned visual thought is preferable to more explicit feedback mechanisms for paths that use construction or hypothesis roles. On MMMU, we compare our readable visual-thought trace against two variants: \emph{latent feedback}, which replaces the $p_{\mathrm{C}}/p_{\mathrm{H}}$ visual-thought step with generated pure visual latents, and \emph{image feedback}, which generates an intermediate image and feeds it back for subsequent reasoning. The comparison is reported both for fixed $p_{\mathrm{C}}/p_{\mathrm{H}}$ execution and for the routed MMMU setting where only the selected $p_{\mathrm{C}}/p_{\mathrm{H}}$ calls are replaced.

The motivation is that both explicit alternatives have practical drawbacks. Full image feedback is expensive, and some intermediate visual thoughts are difficult to synthesize as precise images, so errors in the generated image can hurt later reasoning. Pure visual-latent feedback avoids rendering an image, but it places nonlinguistic visual states inside a text reasoning context, which can break semantic continuity. Aligned visual thought keeps the intermediate step as readable text while using image-derived supervision to shape its hidden representation.

Table~\ref{tab:aligned_visual_thought} shows that explicit feedback is not only slower, but also less accurate. Replacing aligned visual thoughts with newly generated latents drops routed MMMU accuracy by 4.44 points, while feeding back generated intermediate images drops it by 3.44 points. The path-specific results show the same trend on both $p_{\mathrm{C}}$ and $p_{\mathrm{H}}$, where the distinction between visual-thought supervision and explicit visual feedback matters most. In runtime, aligned visual thought reduces per-sample cost by 27.3--30.3\% relative to latent feedback and by 24.4--28.6\% relative to image feedback. These results support the intended design: keeping the intermediate step as text preserves semantic continuity for later reasoning, while image-derived supervision injects visual information into the hidden representation without paying the cost or brittleness of generating and reinserting a concrete visual object.

\begin{center}
\centering
\scriptsize
\setlength{\tabcolsep}{5.0pt}
\renewcommand{\arraystretch}{1.06}
\captionsetup{type=table,hypcap=false}
\caption{MMMU analysis of aligned visual thought versus explicit latent/image feedback. Accuracy is reported on 900 examples. Runtime is average seconds per sample for fixed $p_{\mathrm{C}}$ or $p_{\mathrm{H}}$ execution.}
\label{tab:aligned_visual_thought}
\resizebox{0.86\textwidth}{!}{
\begin{tabular}{lccc|cc}
\toprule
Variant & Routed MMMU & Fixed $p_{\mathrm{C}}$ & Fixed $p_{\mathrm{H}}$ & $p_{\mathrm{C}}$ time & $p_{\mathrm{H}}$ time \\
\midrule
\rowcolor{oursrow}
Aligned visual thought (ours) & 54.44 (490/900) & 52.33 (471/900) & 51.00 (459/900) & 29.05 & 25.37 \\
Latent feedback & 50.00 (450/900) & 48.00 (432/900) & 45.67 (411/900) & 39.98 & 36.41 \\
Image feedback & 51.00 (459/900) & 48.33 (435/900) & 48.22 (434/900) & 38.42 & 35.51 \\
\bottomrule
\end{tabular}}
\end{center}

\section{More Details and Analysis on Planner}
\label{app:planner_details}
\subsection{Planner Training Details}
\label{app:planner_configuration}

We use a compact planner for all routed understanding experiments. The planner takes a path-aware feature vector and produces one score for each path in $\mathcal{P}=\{p_{\mathrm{A}},p_{\mathrm{U}},p_{\mathrm{R}},p_{\mathrm{C}},p_{\mathrm{H}}\}$. The input feature concatenates an image-summary feature with the last-token and mean text features extracted under each candidate path prompt. This yields a 39,424-dimensional feature vector. The planner itself is a two-hidden-layer MLP, where the two hidden layers have width 768 and the output layer has dimension 5.
The five output logits are converted to path-wise scores with a sigmoid. The final route is selected after the query-form calibration step described in Sec.~\ref{sec:experimental_setup}.

Planner supervision is multi-label, since more than one path can solve the same input. We train the planner with weighted binary cross-entropy over the five path labels. This paragraph maps the notation in Sec.~\ref{sec:planner_executor} to the implementation. Samples where all five paths are incorrect are removed from planner training. For each remaining sample, let $n_i=\sum_{p\in\mathcal{P}}r_{i,p}$ be the number of positive paths. The sample weight is $\omega_i=3.0$ when $n_i=1$, $\omega_i=2.0$ when $n_i=2$, and $\omega_i=1.0$ when $n_i\geq3$. The path-level label weight is $\beta_{i,p}=1.3$ only for positive labels outside $p_{\mathrm{A}}$, namely $r_{i,p}=1$ and $p\neq p_{\mathrm{A}}$, and $\beta_{i,p}=1.0$ otherwise. The regularization coefficient $\lambda_{\mathrm{reg}}$ is implemented through AdamW weight decay, set to $5\times 10^{-5}$. We use batch size 256 and learning rate $5\times 10^{-4}$.

\subsection{Planner Training Transfer}
\label{app:planner_training_transfer}

Figure~\ref{fig:planner_dataset_behavior}(c) in the main text shows how held-out planner validation utility transfers to routed MMMU accuracy. Here we additionally report the selected-path distributions of the same five planner checkpoints or configurations. The path distributions reveal a clearer failure mode: lower-utility planners often collapse to a narrow path preference, such as routing 856/900 examples to $p_{\mathrm{R}}$ or 690/900 examples to $p_{\mathrm{U}}$. This supports the role of planner training in learning nontrivial path selection, rather than relying only on hand-crafted query-form calibration.

\begin{center}
  \centering
  \includegraphics[width=.5\textwidth]{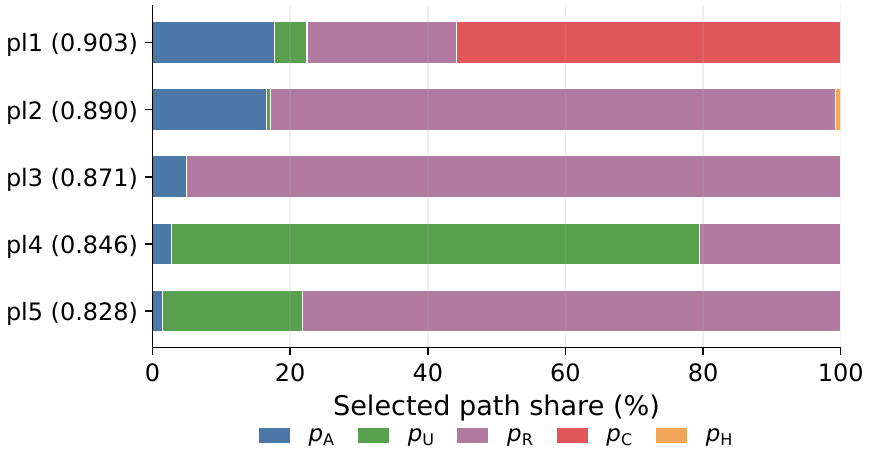}
  \captionsetup{type=figure,hypcap=false}
  \caption{\textbf{Selected path distributions of planner checkpoints.}
  pl1 is the final planner used in our system. Lower-utility planner variants often collapse to a small subset of paths, while the final planner keeps a broader routing pattern.}
  \label{fig:planner_validation_paths}
\end{center}

\subsection{Planner Feature-Space Analysis}
\label{app:planner_feature_space}

To better understand why planner generalization remains difficult, we visualize planner input features from the five understanding benchmarks. Figure~\ref{fig:planner_feature_space} shows that the global feature space is organized much more clearly by dataset/domain than by the path that happens to solve the example. Panel (b) colors each sample by an oracle-correct path label. For examples solved by multiple paths, we randomly sample one correct label for visualization. The resulting labels remain heavily mixed across the global embedding, indicating that limited planner supervision must contend with both domain shift and overlapping path labels.
Panels (c)--(e) visualize local UMAP projections within three representative query-form buckets, denoted Bucket 1--3. Since all displayed samples are correctly solved by our routed model, these local panels color each point by the planner-selected successful path, which is also one of its correct paths. This avoids arbitrary coloring for multi-path examples and shows that bucket-conditioned views are more homogeneous and often expose clearer local path structure than the global path view. The result supports the role of query-form buckets as a lightweight inductive bias: they do not solve path selection by themselves, but they partition the input space before applying learned planner scores, making the routing problem less entangled.

\begin{center}
  \centering
  \includegraphics[width=\textwidth]{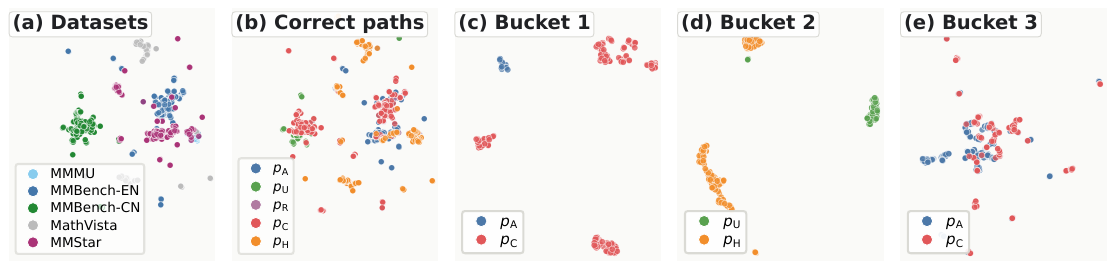}
  \captionsetup{type=figure,hypcap=false}
  \caption{\textbf{Planner feature-space visualization on correctly solved sampled examples.}
  Panels (a)--(b) show the global PCA--UMAP embedding colored by dataset/domain and by one randomly sampled oracle-correct path label for multi-path examples. Panels (c)--(e) show local UMAP projections within three representative buckets, colored by the planner-selected successful path. The global space shows stronger dataset/domain clustering than path separation, while bucket-conditioned views reveal more localized path structure.}
  \label{fig:planner_feature_space}
\end{center}

\section{More Experimental Results}
\label{app:more_experimental_results}

\subsection{Relative Gains of BAGEL-based Post-training Methods}
\label{app:adloop_relative_gain}

Table~\ref{tab:adloop_relative_gain} compares relative improvements over the corresponding BAGEL baseline for BAGEL-based post-training methods. For RecA, UniGame, UniCoT, and Ours, the gains are computed from Table~\ref{tab:understanding_main}. The AD-Loop row is computed from the values reported in its paper.

\begin{center}
\centering
\scriptsize
\setlength{\tabcolsep}{6pt}
\renewcommand{\arraystretch}{1.06}
\captionsetup{type=table,hypcap=false}
\caption{Relative gains over the corresponding BAGEL baseline. RecA, UniGame, UniCoT, and Ours use the BAGEL baseline in Table~\ref{tab:understanding_main}. AD-Loop uses the BAGEL baseline reported in its paper. Missing entries indicate unavailable reports.}
\label{tab:adloop_relative_gain}
\begin{tabular}{lccccc}
\toprule
Method & MMMU & MMB-EN & MMB-CN & MathVista & MMStar \\
\midrule
RecA & +0.8\% & +0.0\% & +0.0\% & -27.9\% & +8.3\% \\
UniGame & +1.0\% & +0.1\% & +0.0\% & +0.8\% & \underline{+8.8\%} \\
UniCoT & +2.3\% & +0.6\% & \underline{+0.1\%} & \underline{+2.0\%} & \textbf{+10.8\%} \\
\midrule
AD-Loop & \underline{+3.6\%} & \underline{+3.1\%} & -- & \textbf{+4.9\%} & +5.5\% \\
\rowcolor{oursrow}
Ours & \textbf{+4.3\%} & \textbf{+4.4\%} & \textbf{+3.2\%} & +0.8\% & +7.7\% \\
\bottomrule
\end{tabular}
\end{center}

\subsection{Path Format Compliance on MMMU}
\label{app:path_format_compliance}

We audit whether the executor follows the requested path template during fixed-path MMMU evaluation. Section format checks whether the output contains the role sections required by the path, such as \texttt{Understanding}, \texttt{Reasoning}, \texttt{Visual}, \texttt{Hypothesis}, and \texttt{Answer}. Pred answer format checks whether the parsed \texttt{pred\_answer} is a legal answer, namely a valid option letter for multiple-choice questions or a nonempty answer for open questions. Strict answer-text format additionally checks whether the raw answer span follows the expected answer surface form.

\begin{center}
\centering
\scriptsize
\setlength{\tabcolsep}{4.8pt}
\renewcommand{\arraystretch}{1.06}
\captionsetup{type=table,hypcap=false}
\caption{Path-format compliance on fixed-path MMMU evaluation. Format checks are high across all paths, showing that the executor usually follows the requested coordination template.}
\label{tab:path_format_compliance}
\resizebox{.95\textwidth}{!}{
\begin{tabular}{lccccc}
\toprule
Path & Section format & Pred. answer format & Strict answer-text format & Acc. & Avg. tokens \\
\midrule
$p_{\mathrm{A}}$ & 900 / 900 = 100.00\% & 900 / 900 = 100.00\% & 900 / 900 = 100.00\% & 51.89 & 1.2 \\
$p_{\mathrm{U}}$ & 899 / 900 = 99.89\% & 900 / 900 = 100.00\% & 899 / 900 = 99.89\% & 50.11 & 74.9 \\
$p_{\mathrm{R}}$ & 880 / 900 = 97.78\% & 900 / 900 = 100.00\% & 875 / 900 = 97.22\% & 52.22 & 231.4 \\
$p_{\mathrm{C}}$ & 886 / 900 = 98.44\% & 900 / 900 = 100.00\% & 886 / 900 = 98.44\% & 52.33 & 291.5 \\
$p_{\mathrm{H}}$ & 886 / 900 = 98.44\% & 900 / 900 = 100.00\% & 887 / 900 = 98.56\% & 51.00 & 295.7 \\
\bottomrule
\end{tabular}}
\end{center}

The executor follows the requested section structure for nearly all examples and always produces a parsable final answer. Thus, downstream failures are unlikely to be dominated by simple template-following errors. Together with the planner analyses in the main text, this suggests that the larger remaining bottleneck is selecting the right path for each query rather than making the executor emit the requested path format.

\subsection{Different Backbone Experiments}
\label{app:different_backbone}

We further instantiate the same coordination path space with a smaller Harmon-1.5B~\citep{wu2025harmonizing} backbone. This experiment is intended to test whether path complementarity is tied only to the main BAGEL backbone. Table~\ref{tab:harmon_path_oracle} shows that single-path performance remains close to raw Harmon performance, but the five-path oracle reaches 67.78\% on MMMU. The gain over raw Harmon is 33.45 points, indicating that the path space still exposes substantial complementary behavior under a different and smaller executor.

\begin{center}
\centering
\scriptsize
\setlength{\tabcolsep}{6pt}
\renewcommand{\arraystretch}{1.06}
\captionsetup{type=table,hypcap=false}
\caption{MMMU path complementarity with a Harmon-1.5B backbone. The oracle selects the correct answer whenever any of the five path-conditioned executions is correct.}
\label{tab:harmon_path_oracle}
\begin{tabular}{lccccccc}
\toprule
Metric & Raw & $p_{\mathrm{A}}$ & $p_{\mathrm{U}}$ & $p_{\mathrm{R}}$ & $p_{\mathrm{C}}$ & $p_{\mathrm{H}}$ & Oracle \\
\midrule
Correct / Total & 309 / 900 & 302 / 900 & 296 / 900 & 272 / 900 & 289 / 900 & 252 / 900 & 610 / 900 \\
Acc. & 34.33 & 33.56 & 32.89 & 30.22 & 32.11 & 28.00 & 67.78 \\
\bottomrule
\end{tabular}
\end{center}

Table~\ref{tab:harmon_routing} reports deployable routed results on five understanding benchmarks. The routed system improves over raw Harmon on four of the five datasets, but the gains are much smaller than the oracle gap in Table~\ref{tab:harmon_path_oracle}. This suggests that the coordination paths remain useful across backbones, while converting path complementarity into reliable routed gains becomes harder with a weaker executor. 

\begin{center}
\centering
\scriptsize
\setlength{\tabcolsep}{6pt}
\renewcommand{\arraystretch}{1.06}
\captionsetup{type=table,hypcap=false}
\caption{Routed results with the Harmon-1.5B backbone. Scores are accuracies in percent.}
\label{tab:harmon_routing}
\begin{tabular}{lcccccc}
\toprule
Method & MMMU & MathVista & MMStar & MMB-EN & MMB-CN & Average \\
\midrule
Raw & 34.33 & 24.50 & 37.47 & 62.72 & 54.65 & 42.73 \\
Ours & 34.78 & 28.70 & 38.00 & 63.06 & 54.65 & 43.84 \\
\bottomrule
\end{tabular}
\end{center}

\subsection{Additional Generation Results}
\label{app:generation_full}

Tables~\ref{tab:appendix_geneval_full} and~\ref{tab:appendix_wise_full} expand the generation results summarized in the main text. We report the category-level GenEval and WISE scores for the same models as Table~\ref{tab:understanding_main}, so that the overall generation numbers can be traced back to their underlying subcategories.

\begin{center}
\centering
\scriptsize
\setlength{\tabcolsep}{3.4pt}
\renewcommand{\arraystretch}{1.04}
\captionsetup{type=table,hypcap=false}
\caption{Complete GenEval results. Scores are accuracies in percent.}
\label{tab:appendix_geneval_full}
\resizebox{.8\textwidth}{!}{
\begin{tabular}{lccccccc}
\toprule
Method & Single object & Two object & Counting & Colors & Position & Color attr. & Overall \\
\midrule
BAGEL & 99.38 & 94.19 & 78.75 & 87.77 & 51.00 & 61.75 & 78.81 \\
BLIP3-o & 98.12 & 93.18 & 73.44 & 86.17 & 72.75 & 64.50 & 81.36 \\
Show-o2 & 97.81 & 71.46 & 48.75 & 78.46 & 20.00 & 42.75 & 59.87 \\
Show-o2 (1.5B) & 96.88 & 64.39 & 46.88 & 76.06 & 16.75 & 32.00 & 55.49 \\
Janus-Pro & 97.81 & 86.62 & 57.50 & 89.36 & 76.00 & 66.25 & 78.92 \\
Janus & 85.62 & 37.63 & 18.75 & 53.46 & 17.50 & 27.25 & 40.04 \\
JanusFlow & 94.25 & 46.06 & 27.75 & 74.68 & 32.20 & 25.00 & 49.99 \\
OmniGen2 & 99.69 & 93.94 & 68.75 & 88.03 & 53.25 & 67.50 & 78.53 \\
TokenFlow & 97.19 & 59.60 & 37.81 & 86.17 & 17.25 & 15.25 & 52.21 \\
Emu3 & 94.69 & 55.81 & 30.00 & 76.06 & 8.50 & 9.50 & 45.76 \\
DeepGen & 98.75 & 98.99 & 81.25 & 92.55 & 75.00 & 73.00 & 86.59 \\
MMaDA & 89.06 & 49.75 & 31.25 & 73.67 & 12.50 & 20.50 & 46.12 \\
Emu3.5 & 100.00 & 93.94 & 49.06 & 91.49 & 85.50 & 71.00 & 81.83 \\
Ovis-U1 & 100.00 & 94.95 & 93.75 & 93.62 & 80.00 & 78.00 & 90.05 \\
\midrule
IRG & 98.44 & 87.37 & 70.31 & 78.72 & 40.50 & 57.00 & 72.06 \\
RecA & 99.38 & 94.44 & 79.38 & 89.10 & 61.75 & 74.25 & 83.05 \\
UniCoT & 98.44 & 91.92 & 80.94 & 86.44 & 47.75 & 62.00 & 77.91 \\
UniGame & 98.44 & 95.96 & 81.25 & 93.62 & 72.00 & 75.75 & 86.17 \\
\rowcolor{oursrow}
Ours & 98.00 & 95.00 & 77.00 & 86.00 & 54.00 & 69.00 & 80.00 \\
\bottomrule
\end{tabular}}
\end{center}

\begin{center}
\centering
\scriptsize
\setlength{\tabcolsep}{3.4pt}
\renewcommand{\arraystretch}{1.04}
\captionsetup{type=table,hypcap=false}
\caption{Complete WISE results. Scores are WiScore.}
\label{tab:appendix_wise_full}
\resizebox{.8\textwidth}{!}{
\begin{tabular}{lccccccc}
\toprule
Method & Culture & Time & Space & Biology & Physics & Chemistry & Overall \\
\midrule
BAGEL & 0.3883 & 0.4386 & 0.4714 & 0.3620 & 0.4205 & 0.2940 & 0.3989 \\
BLIP3-o & 0.4028 & 0.4186 & 0.5259 & 0.4025 & 0.4255 & 0.3000 & 0.4138 \\
Show-o2 & 0.3641 & 0.3497 & 0.4519 & 0.3455 & 0.3690 & 0.2390 & 0.3595 \\
Show-o2 (1.5B) & 0.3111 & 0.3563 & 0.4357 & 0.3150 & 0.3840 & 0.2310 & 0.3349 \\
Show-o & 0.2865 & 0.3225 & 0.4132 & 0.2750 & 0.3300 & 0.1980 & 0.3037 \\
Janus-Pro & 0.3616 & 0.3853 & 0.4789 & 0.3605 & 0.4745 & 0.2485 & 0.3811 \\
Janus & 0.2080 & 0.2707 & 0.3508 & 0.1705 & 0.1910 & 0.1095 & 0.2222 \\
JanusFlow & 0.2731 & 0.3222 & 0.3947 & 0.3215 & 0.2860 & 0.1905 & 0.2954 \\
OmniGen2 & 0.4180 & 0.4042 & 0.4887 & 0.3635 & 0.3875 & 0.2810 & 0.4029 \\
TokenFlow & 0.3253 & 0.3626 & 0.3357 & 0.2915 & 0.2605 & 0.1510 & 0.3056 \\
Emu3 & 0.3463 & 0.3482 & 0.3711 & 0.3310 & 0.3685 & 0.2130 & 0.3373 \\
DeepGen & 0.5989 & 0.4955 & 0.6102 & 0.4765 & 0.5515 & 0.4080 & 0.5470 \\
Emu3.5 & 0.7001 & 0.5683 & 0.6944 & 0.6435 & 0.6085 & 0.4060 & 0.6331 \\
MMaDA & 0.6502 & 0.6814 & 0.7492 & 0.6620 & 0.7420 & 0.4205 & 0.6560 \\
Ovis-U1 & 0.3643 & 0.3991 & 0.4831 & 0.3405 & 0.4090 & 0.2390 & 0.3755 \\
\midrule
IRG & 0.3674 & 0.4081 & 0.4650 & 0.3575 & 0.4495 & 0.2655 & 0.3842 \\
RecA & 0.4035 & 0.4147 & 0.5432 & 0.3985 & 0.4630 & 0.3340 & 0.4225 \\
UniCoT & 0.3998 & 0.4183 & 0.4797 & 0.3405 & 0.4550 & 0.3060 & 0.4037 \\
UniGame & 0.3956 & 0.4138 & 0.4876 & 0.3590 & 0.4355 & 0.3155 & 0.4032 \\
\rowcolor{oursrow}
Ours & 0.3905 & 0.4287 & 0.4962 & 0.4360 & 0.4370 & 0.2890 & 0.4100 \\
\bottomrule
\end{tabular}}
\end{center}

\section{Executor Training Details}
\label{app:executor_training_details}

This appendix summarizes the final staged LoRA executor used in the main experiments. We initialize from BAGEL-7B-MoT, freeze the base model, and train language-model LoRA adapters together with the lightweight projection head used for aligned visual-thought supervision. The LoRA adapter uses rank 16, alpha 32, and dropout 0.05. We report only the hyperparameters needed to interpret the method-level objectives in Sec.~\ref{sec:method}.

Executor training is organized into four staged splits that activate different links of the path: text-only understanding (S1), understanding with aligned visual thoughts (S2), final image generation without aligned visual-thought supervision (S3), and final image generation with construction or hypothesis supervision (S4).
The four stages form a strict checkpoint chain: S1 best is used to initialize S2, S2 best initializes S3, and S3 best initializes S4. All stages use batch size 1, bf16 training, weight decay 0.01, evaluation every 200 steps, and GH200 GPUs. S1 and S2 are trained for one epoch with maximum sequence lengths 3072 and 6144, respectively. S3 and S4 are trained for 3000 steps. Best-checkpoint selection enforces no large degradation in format accuracy, with tolerance 0.01 for S1 and 0.03 for S2--S4. S1 uses early-stop patience 6 with a 1200-step warmup before stopping can trigger, while S2--S4 use patience 8 with a 1600-step warmup. For S3 and S4 generation-side validation, evaluation allows up to 2048 new tokens.

The loss weights in Table~\ref{tab:appendix_executor_chain} correspond directly to the symbols in Sec.~\ref{sec:planner_executor}. In S1 and S2, $w_t$ is implemented as a token-level \texttt{loss\_weights} vector: prompt and context tokens have weight 0, intermediate role tokens use the ``thought CE'' value, and final \texttt{Answer} tokens use the ``answer CE'' value. The resulting normalized $\mathcal{L}_{\mathrm{text}}$ is used with $\lambda_{\mathrm{text}}=1$. In S3 and S4, the native image-answer trainer uses unit token weights for supervised target text and controls the text term with the global CE coefficient, so the table reports this value as $\lambda_{\mathrm{text}}$. The MSE and visual columns are $\lambda_{\mathrm{mse}}$ and $\lambda_{\mathrm{vis}}$, with absent losses set to 0.

\begin{algorithm}[t]
\caption{Staged executor training pipeline.}
\label{alg:training_pipeline}
\footnotesize
\begin{algorithmic}[1]
\Require Base UMM $M_0$, staged role-aligned data $\mathcal{D}_{1:4}$, latent cache $\mathcal{V}$
\Ensure Path-conditioned executor $\mathcal{E}_{\theta}$
\State Initialize the executor from $M_0$
\State Attach language-model LoRA adapters and projection head $g_{\phi}$, then freeze the base UMM
\State Set $C_0 \leftarrow M_0$ with trainable LoRA parameters and $g_{\phi}$
\For{$s=1$ to $4$}
    \State Load stage data $\mathcal{D}_s$ and initialize from checkpoint $C_{s-1}$
    \For{minibatch $B \subset \mathcal{D}_s$}
        \State Convert each example to tagged role segments according to its path $p$
        \State Compute the role-weighted text loss $\mathcal{L}_{\mathrm{text}}$ on supervised text tokens
        \If{$B$ contains \texttt{Visual} or \texttt{Hypothesis} spans}
            \State Retrieve the corresponding reference visual summaries $\{v_j\}$ from $\mathcal{V}$
            \State Pool hidden states $\{\bar{h}_j\}$ over the tagged visual-thought spans
            \State Compute $\mathcal{L}_{\mathrm{vis}}=\frac{1}{J}\sum_j\|g_{\phi}(\bar{h}_j)-v_j\|_2^2$
        \Else
            \State Set $\mathcal{L}_{\mathrm{vis}} \leftarrow 0$
        \EndIf
        \If{$B$ contains final image answers}
            \State Compute BAGEL's final image-latent reconstruction loss $\mathcal{L}_{\mathrm{latent}}$
        \Else
            \State Set $\mathcal{L}_{\mathrm{latent}} \leftarrow 0$
        \EndIf
        \State Update trainable parameters with $\lambda_{\mathrm{text}}\mathcal{L}_{\mathrm{text}}+\lambda_{\mathrm{mse}}\mathcal{L}_{\mathrm{latent}}+\lambda_{\mathrm{vis}}\mathcal{L}_{\mathrm{vis}}$
    \EndFor
    \State Select the best checkpoint $C_s$ using the stage validation metric and format-preservation gate
\EndFor
\State Set $\mathcal{E}_{\theta} \leftarrow C_4$
\State \Return $\mathcal{E}_{\theta}$
\end{algorithmic}
\end{algorithm}

\begin{center}
\centering
\scriptsize
\setlength{\tabcolsep}{3pt}
\renewcommand{\arraystretch}{1.12}
\captionsetup{type=table,hypcap=false}
\caption{Staged executor training hyperparameters. The final column gives the code-level settings for the token weights $w_t$ and objective coefficients $\lambda_{\mathrm{text}}$, $\lambda_{\mathrm{mse}}$, and $\lambda_{\mathrm{vis}}$ in Sec.~\ref{sec:planner_executor}.}
\label{tab:appendix_executor_chain}
\resizebox{\textwidth}{!}{
\begin{tabular}{p{0.08\textwidth}p{0.23\textwidth}p{0.14\textwidth}p{0.15\textwidth}p{0.11\textwidth}p{0.29\textwidth}}
\toprule
Stage & Role split & Initialization & Hardware & LR & $w_t$ and $\lambda$ settings \\
\midrule
S1 & Text understanding, $p_{\mathrm{U}}/p_{\mathrm{R}}$ & BAGEL init & 2 GH200 GPUs & $3{\times}10^{-6}$ & $w_t{=}0.5$ for thought tokens, $w_t{=}4.0$ for answer tokens, $\lambda_{\mathrm{text}}{=}1$, $\lambda_{\mathrm{mse}}{=}0$, $\lambda_{\mathrm{vis}}{=}0$ \\
S2 & Visual-thought, $p_{\mathrm{C}}/p_{\mathrm{H}}$ & S1 best & 4 GH200 GPUs & $4{\times}10^{-6}$ & $w_t{=}0.25$ for thought tokens, $w_t{=}6.0$ for answer tokens, $\lambda_{\mathrm{text}}{=}1$, $\lambda_{\mathrm{mse}}{=}0$, $\lambda_{\mathrm{vis}}{=}0.05$ \\
S3 & Plain image answer & S2 best & 4 GH200 GPUs & $3{\times}10^{-6}$ & unit target-token weights, $\lambda_{\mathrm{text}}{=}2.0$, $\lambda_{\mathrm{mse}}{=}0.3$, $\lambda_{\mathrm{vis}}{=}0$ \\
S4 & Image answer + visual & S3 best & 4 GH200 GPUs & $3{\times}10^{-6}$ & unit target-token weights, $\lambda_{\mathrm{text}}{=}2.0$, $\lambda_{\mathrm{mse}}{=}0.3$, $\lambda_{\mathrm{vis}}{=}0.05$ \\
\bottomrule
\end{tabular}}
\end{center}

For staged executor diagnostics, answer accuracy reports final-answer correctness on understanding splits. Format accuracy measures whether the output follows the expected role structure, so it is a sanity check for path execution rather than evidence of correct reasoning. CE denotes the weighted language-modeling loss on supervised text tokens. Visual denotes the alignment loss between textual visual-thought hidden states and image-derived visual summaries. MSE denotes BAGEL's final image-latent reconstruction loss for generation-side stages.

\subsection{Executor Data Statistics}
\label{app:executor_data_stats}

Table~\ref{tab:appendix_executor_data_stats} reports the split sizes used by the staged executor run after filtering and stage assignment. These counts correspond to the four validation splits used in Table~\ref{tab:executor_diagnostics}. They describe the executor-stage data rather than the additional path-outcome runs used to build planner supervision.

\begin{center}
\centering
\small
\setlength{\tabcolsep}{5pt}
\renewcommand{\arraystretch}{1.08}
\captionsetup{type=table,hypcap=false}
\caption{Staged executor split statistics.}
\label{tab:appendix_executor_data_stats}
\begin{tabular}{llrrr}
\toprule
Stage & Split name & Train & Val & Total \\
\midrule
S1 & \texttt{understanding\_text} & 12,733 & 164 & 12,897 \\
S2 & \texttt{understanding\_visual} & 5,232 & 68 & 5,300 \\
S3 & \texttt{image\_answer\_plain} & 5,380 & 92 & 5,472 \\
S4 & \texttt{image\_answer\_visual} & 6,282 & 115 & 6,397 \\
\midrule
Total & -- & 29,627 & 439 & 30,066 \\
\bottomrule
\end{tabular}
\end{center}

\subsection{Executor Training Diagnostics}
\label{app:executor_diagnostics}

Before the planner is evaluated on downstream benchmarks, we compare several executor training strategies on the four staged validation splits. This diagnostic separates superficial format following from useful path execution. A valid trajectory should both obey the role structure and improve the task-relevant objective. Figure~\ref{fig:executor_training_curves} shows the smoothed training losses of the final staged LoRA executor, and Table~\ref{tab:executor_diagnostics} reports the raw BAGEL executor and three finetuning variants on held-out validation splits. The \emph{staged LoRA} setting trains one stage at a time and passes the previous adapter forward. This is the executor used in our final experiments. \emph{Partial-SFT} unfreezes a subset of the language model instead of using PEFT adapters~\citep{peft}, while \emph{multitask Partial-SFT} mixes all stages in a single training run.

\begin{center}
  \centering
  \includegraphics[width=.9\textwidth]{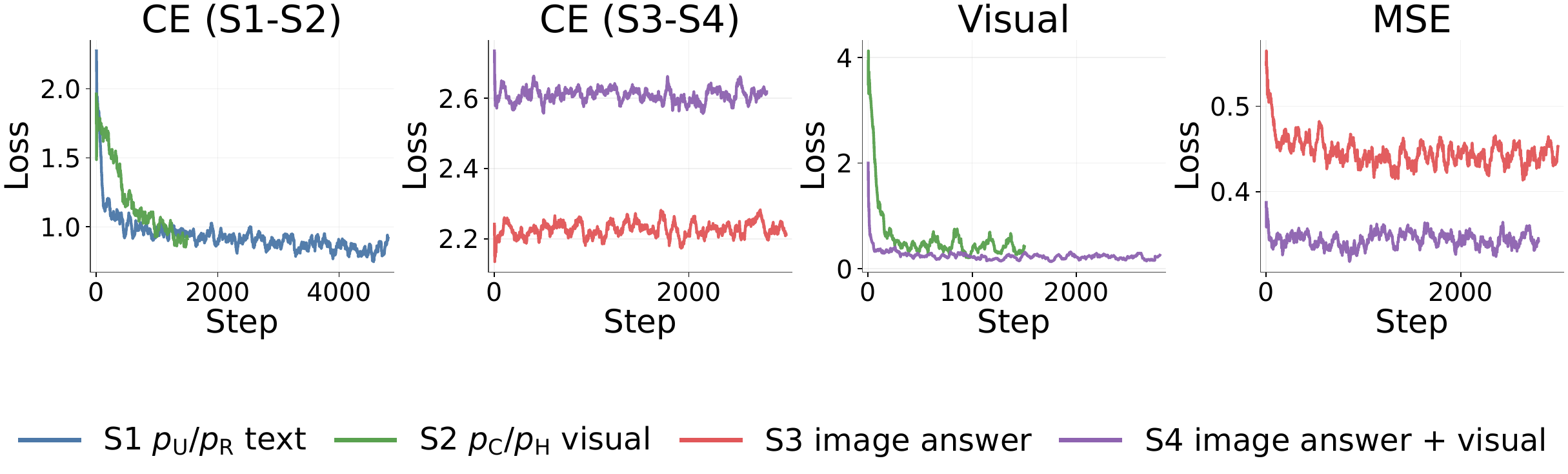}
  \captionsetup{type=figure,hypcap=false}
  \caption{\textbf{Smoothed training losses for the staged LoRA executor.}
  Panels are arranged in one row and use the same stage names as Table~\ref{tab:executor_diagnostics}. CE decreases clearly on S1/S2, while visual and MSE losses show that aligned visual-thought and image-latent objectives are actively optimized in the stages where they are enabled.}
  \label{fig:executor_training_curves}
\end{center}

\begin{center}
\centering
\scriptsize
\captionsetup{type=table,hypcap=false}
\caption{Executor training diagnostics on the four staged validation splits. Acc. and Format are higher-is-better, while CE, Visual, and MSE are lower-is-better. Although raw BAGEL often preserves the requested format, its low answer accuracy and weak intermediate/final losses indicate that format compliance alone does not provide meaningful path execution. The final system uses staged LoRA.}
\label{tab:executor_diagnostics}
\resizebox{.9\textwidth}{!}{
\begin{tabular}{llccccc}
\toprule
Stage & Method & Acc. & Format & CE & Visual & MSE \\
\midrule
S1 $p_{\mathrm{U}}/p_{\mathrm{R}}$ text & BAGEL & 0.2439 & 1.0000 & 1.7557 & -- & -- \\
\rowcolor{oursrow}
 & Staged LoRA & 0.8125 & 1.0000 & 0.9775 & -- & -- \\
 & Partial-SFT & 0.3476 & 1.0000 & 1.4324 & -- & -- \\
 & Multitask Partial-SFT & 0.4531 & 1.0000 & 1.9636 & -- & -- \\
\midrule
S2 $p_{\mathrm{C}}/p_{\mathrm{H}}$ visual & BAGEL & 0.2353 & 1.0000 & 2.8750 & 4.0301 & -- \\
\rowcolor{oursrow}
 & Staged LoRA & 0.3253 & 0.7390 & 0.5384 & 0.3523 & -- \\
 & Partial-SFT & 0.3235 & 1.0000 & 1.8601 & 3.4149 & -- \\
 & Multitask Partial-SFT & 0.2500 & 0.9844 & 2.0606 & 6.4729 & -- \\
\midrule
S3 image answer & BAGEL & -- & 1.0000 & 2.3748 & -- & 0.5410 \\
\rowcolor{oursrow}
 & Staged LoRA & -- & 1.0000 & 2.2334 & -- & 0.4510 \\
 & Partial-SFT & -- & 1.0000 & 1.7335 & -- & 0.5320 \\
 & Multitask Partial-SFT & -- & 1.0000 & 1.7537 & -- & 0.5376 \\
\midrule
S4 image answer + visual & BAGEL & -- & 0.9478 & 2.7529 & 1.4585 & 0.4560 \\
\rowcolor{oursrow}
 & Staged LoRA & -- & 0.9688 & 2.6233 & 0.2718 & 0.3446 \\
 & Partial-SFT & -- & 1.0000 & 2.6146 & 0.6878 & 0.4453 \\
 & Multitask Partial-SFT & -- & 1.0000 & 2.6545 & 1.0655 & 0.4535 \\
\bottomrule
\end{tabular}}
\end{center}

Several observations guide the final experimental choice. First, the raw BAGEL backbone can often preserve the requested output structure, with near-perfect format accuracy in several splits. However, its S1/S2 answer accuracy remains low and its visual-summary loss is high, indicating that the pretrained model is mostly following the surface template rather than executing the intended intermediate reasoning or visual-grounding operations. Second, format accuracy alone is not sufficient: Partial-SFT and multitasking often maintain perfect formatting while improving task accuracy only weakly. The downstream format audit in Appendix~\ref{app:path_format_compliance} further confirms that the final executor can follow path templates on MMMU. Finally, staged LoRA provides the best balance between stage-level learning, format preservation, checkpoint size, and downstream path behavior, so we use it for the executor in the main experiments.

\section{Path Prompt Templates}
\label{app:path_prompts}

This section lists the execution-time prompt wrappers used for path-conditioned evaluation. For $p_{\mathrm{A}}$, the executor receives the benchmark query directly. For $p_{\mathrm{U}}$, $p_{\mathrm{R}}$, $p_{\mathrm{C}}$, and $p_{\mathrm{H}}$, the template below is prepended and the benchmark query is appended after \texttt{Query:}. These prompts are used after the planner has selected a path. They specify the executor's role order and output format rather than replacing the learned path selector.

\newtcblisting{promptlisting}[2][]{
  enhanced,
  breakable,
  listing only,
  listing engine=listings,
  colback=gray!3,
  colframe=black!35,
  boxrule=0.45pt,
  arc=1pt,
  left=4pt,
  right=4pt,
  top=4pt,
  bottom=4pt,
  title={#2},
  fonttitle=\bfseries\small,
  listing options={
    basicstyle=\ttfamily\scriptsize,
    breaklines=true,
    columns=fullflexible,
    keepspaces=true
  },
  #1
}

\begin{promptlisting}{Prompt for $p_{\mathrm{A}}$}
<benchmark query>
\end{promptlisting}

\begin{promptlisting}{Prompt for $p_{\mathrm{U}}$}
You are a multimodal reasoning model. For this task, you must follow this EXACT reasoning path:

Understanding -> Answer

Where:
- Understanding = Text Understanding: give exactly one sentence that directly describes the visible input and the task, without inference
- Answer = Final Answer: answer directly from perception only

Output format:

Understanding:
(Exactly one sentence directly describing the visible input and the task)

Answer:
(Direct answer based only on what is explicitly visible or stated)

Now solve the following query using ONLY this path (Understanding -> Answer):

Query:
<benchmark query>
\end{promptlisting}

\begin{promptlisting}{Prompt for $p_{\mathrm{R}}$}
You are a multimodal reasoning model. For this task, you must follow this EXACT reasoning path:

Understanding -> Reasoning -> Answer

Where:
- Understanding = Text Understanding: give exactly one sentence that directly describes the visible input and the task, without inference
- Reasoning = Text Reasoning: give a short textual reasoning chain using only the provided information
- Answer = Final Answer: answer directly and concisely

Output format:

Understanding:
(Exactly one sentence directly describing the visible input and the task)

Reasoning:
(A short textual reasoning chain)

Answer:
(Final answer)

Now solve the following query using ONLY this path (Understanding -> Reasoning -> Answer):

Query:
<benchmark query>
\end{promptlisting}

\begin{promptlisting}{Prompt for $p_{\mathrm{C}}$}
You are a multimodal reasoning model. For this task, you must follow this EXACT reasoning path:

Understanding -> Reasoning -> Visual -> Reasoning -> Answer

Where:
- Understanding = Text Understanding: give exactly one sentence that directly describes the visible input and the task, without inference
- Reasoning = Text Reasoning: give a short textual reasoning chain using only the provided information
- Visual = Visual Reasoning: describe one intermediate visual construct or visual step
- Answer = Final Answer: answer directly and concisely

Output format:

Understanding:
(Exactly one sentence directly describing the visible input and the task)

Reasoning:
(A short textual reasoning chain)

Visual:
(One intermediate visual construct or visual step)

Reasoning:
(A short textual reasoning chain that uses the visual step)

Answer:
(Final answer)

Now solve the following query using ONLY this path (Understanding -> Reasoning -> Visual -> Reasoning -> Answer):

Query:
<benchmark query>
\end{promptlisting}

\begin{promptlisting}{Prompt for $p_{\mathrm{H}}$}
You are a multimodal reasoning model. For this task, you must follow this EXACT reasoning path:

Understanding -> Reasoning -> Hypothesis -> Reasoning -> Answer

Where:
- Understanding = Text Understanding: give exactly one sentence that directly describes the visible input and the task, without inference
- Reasoning = Text Reasoning: give a short textual reasoning chain using only the provided information
- Hypothesis = Visual Hypothesis: describe candidate visual states or hypotheses before answering
- Answer = Final Answer: answer directly and concisely

Output format:

Understanding:
(Exactly one sentence directly describing the visible input and the task)

Reasoning:
(A short textual reasoning chain)

Hypothesis:
(Candidate visual states or hypotheses)

Reasoning:
(A short textual reasoning chain that uses the visual hypotheses)

Answer:
(Final answer)

Now solve the following query using ONLY this path (Understanding -> Reasoning -> Hypothesis -> Reasoning -> Answer):

Query:
<benchmark query>
\end{promptlisting}

\section{Additional Qualitative Examples}
\label{app:additional_qualitative_examples}

We provide additional MMMU cases from the full five-path evaluation. In each case, only the highlighted path among $\{p_{\mathrm{A}},p_{\mathrm{U}},p_{\mathrm{R}},p_{\mathrm{C}},p_{\mathrm{H}}\}$ gives the correct answer.

\begin{pathcase}[borderline west={1.3pt}{0pt}{black!55}]{$p_{\mathrm{A}}$: direct answer}
\caseitem{Question}{What are the yellow spheres connected to the roots of a potato? Options: A. Sclerotia, B. Cysts, C. Eggs, D. Galls.}
\casepred{\okpred{B}}{A}{A}{A}{A}{B}
\casetrace{\tracehead{Direct output}{B}}
\caseitem{Why this path}{The visual information needed by the answer is already described by the query itself: yellow spheres are connected to potato roots. Additional decomposition is unnecessary and can move the model toward plausible but wrong distractors.}
\end{pathcase}

\begin{pathcase}[borderline west={1.3pt}{0pt}{easy_color}, colframe=easy_color!35!black]{$p_{\mathrm{U}}$: visual understanding only}
\caseitem{Question}{A velocity--time graph asks when the object again passes through its initial position. Options: A. 1 s, B. Between 1 and 2 s, C. 2 s, D. Between 2 and 3 s.}
\casepred{C}{\okpred{B}}{C}{C}{C}{B}
\casetrace{\tracehead{Understanding}{The velocity curve starts at $(0,0)$, dips below zero before returning to $0$ at $t=1$, and then rises above zero toward $2$ by $t=2$. The return to the initial position is indicated between $1$ and $2$ s.}
\tracehead{Answer}{B}}
\caseitem{Why this path}{The decisive evidence is the numerical structure of the plotted curve. Explicit perception preserves the visual anchors needed to choose the interval, while additional roles lead the model toward incorrect alternatives.}
\end{pathcase}

\begin{pathcase}[borderline west={1.3pt}{0pt}{medium_color}, colframe=medium_color!35!black]{$p_{\mathrm{R}}$: textual reasoning}
\caseitem{Question}{Given a chemical structure with a central Pt atom bonded to two ammine and two chlorine ligands, identify the untrue statement. Options: A. Chlorine groups are displaced before activation, B. The agent causes intrastrand DNA cross-linking, C. Adenine--thymine base pairing is disrupted, D. The compound acts as a metallating agent.}
\casepred{A}{A}{\okpred{C}}{A}{A}{C}
\casetrace{\tracehead{Understanding}{The image shows a platinum complex with a central Pt atom bonded to two ammonia groups and two chlorine atoms.}
\tracehead{Reasoning}{The structure is cisplatin. It acts as a metallating agent and leads to intrastrand DNA cross-linking. Base pairing between adenine and thymine is not disrupted by the agent.}
\tracehead{Answer}{C}}
\caseitem{Why this path}{The visual input must be interpreted, but the answer is determined by textual biochemical reasoning rather than by constructing an aligned visual thought.}
\end{pathcase}

\begin{pathcase}[borderline west={1.3pt}{0pt}{hard_color}, colframe=hard_color!45!black]{$p_{\mathrm{H}}$: visual hypotheses}
\caseitem{Question}{Which graph shows how dissipated power $P$ varies with current $I$ for an Ohmic component? Options: A. upward-curving from the origin, B. decreasing linear, C. increasing linear, D. decreasing curved.}
\casepred{C}{C}{C}{C}{\okpred{A}}{A}
\casetrace{\tracehead{Understanding}{Four graphs labeled A, B, C, and D are shown, each plotting $P$ against $I$.}
\tracehead{Reasoning}{For an Ohmic component, $V=IR$ and $P=I^2R$, so power is proportional to the square of current.}
\tracehead{Hypothesis}{Candidate visual hypotheses correspond to the four graph shapes: A is upward-curving from the origin, B is decreasing linear, C is increasing linear, and D is decreasing curved.}
\tracehead{Reasoning}{Graph A matches the expected upward-curving relation, while the linear and decreasing alternatives do not.}
\tracehead{Answer}{A}}
\caseitem{Why this path}{The difficulty is not only reading the figure, but choosing among plausible visual hypotheses. Explicit hypothesis comparison prevents the model from defaulting to the tempting linear option.}
\end{pathcase}

\section{Role-Aligned Trajectory Examples}
\label{app:trajectory_examples}

We show one representative supervised trajectory for each coordination path beyond $p_{\mathrm{A}}$. The examples are drawn from the staged executor data. Role contents are lightly shortened for readability while preserving the supervised role order. For $p_{\mathrm{C}}$ and $p_{\mathrm{H}}$, the displayed visual references are used to supervise aligned visual thoughts during training. At inference time, the executor emits the tagged textual trace rather than inserting these images back into the context.

\newtcolorbox{trajcase}[2][]{
  enhanced,
  breakable,
  colback=white,
  colframe=black!14,
  colbacktitle=headergray,
  coltitle=black,
  title={#2},
  fonttitle=\bfseries\small,
  fontupper=\scriptsize,
  boxrule=0.45pt,
  arc=1.5pt,
  left=5pt,
  right=5pt,
  top=4pt,
  bottom=4pt,
  before skip=6pt,
  after skip=8pt,
  #1
}
\newcommand{\trajfield}[2]{\textcolor{black!62}{\textsc{#1.}}~#2\par\smallskip}
\newcommand{\trajrole}[2]{\textbf{#1:}~#2\par}
\newcommand{\trajimglabel}[1]{{\footnotesize\textcolor{black!58}{#1}}\par}

\begin{trajcase}[borderline west={1.3pt}{0pt}{easy_color}]{$p_{\mathrm{U}}$: Understanding $\rightarrow$ Answer}
\begin{minipage}[t]{0.31\textwidth}
\vspace{0pt}
\centering
\includegraphics[width=\linewidth]{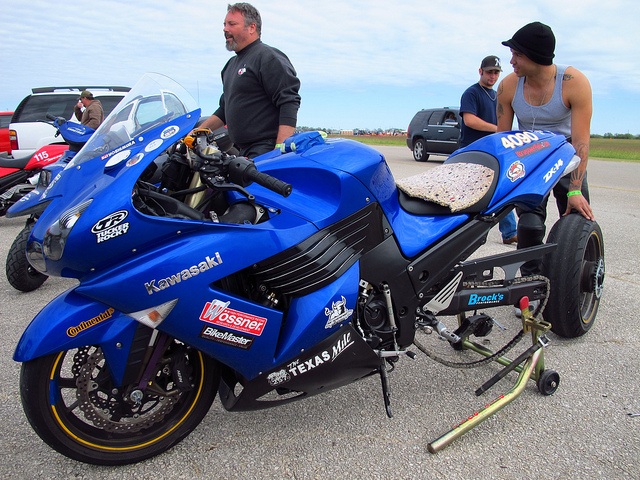}
\trajimglabel{Input image}
\end{minipage}
\hfill
\begin{minipage}[t]{0.64\textwidth}
\vspace{0pt}
\trajfield{Query}{What brand is the bike?}
\trajrole{Understanding}{A blue Kawasaki sport motorcycle with racing decals is parked on pavement, with people standing nearby and other vehicles in the background.}
\trajrole{Answer}{kawasaki}
\end{minipage}
\end{trajcase}

\begin{trajcase}[borderline west={1.3pt}{0pt}{medium_color}, colframe=medium_color!30!black]{$p_{\mathrm{R}}$: Understanding $\rightarrow$ Reasoning $\rightarrow$ Answer}
\begin{minipage}[t]{0.31\textwidth}
\vspace{0pt}
\centering
\includegraphics[width=\linewidth]{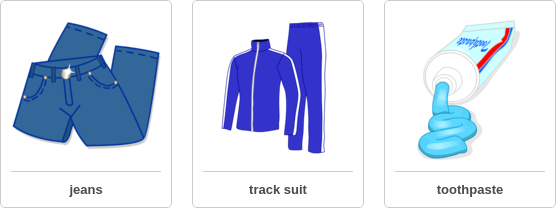}
\trajimglabel{Input image}
\end{minipage}
\hfill
\begin{minipage}[t]{0.64\textwidth}
\vspace{0pt}
\trajfield{Query}{Which property do these three objects have in common? Choices: (A) fragile, (B) blue, (C) sticky.}
\trajrole{Understanding}{The image shows jeans, a track suit, and toothpaste, all containing visible blue regions.}
\trajrole{Reasoning}{Fragile and sticky are not shared by all three objects. Blue is the shared visual property.}
\trajrole{Answer}{blue}
\end{minipage}
\end{trajcase}

\begin{trajcase}[borderline west={1.3pt}{0pt}{traj_color}, colframe=traj_color!35!black]{$p_{\mathrm{C}}$: Understanding $\rightarrow$ Reasoning $\rightarrow$ Visual $\rightarrow$ Reasoning $\rightarrow$ Answer}
\begin{minipage}[t]{0.38\textwidth}
\vspace{0pt}
\centering
\includegraphics[width=\linewidth]{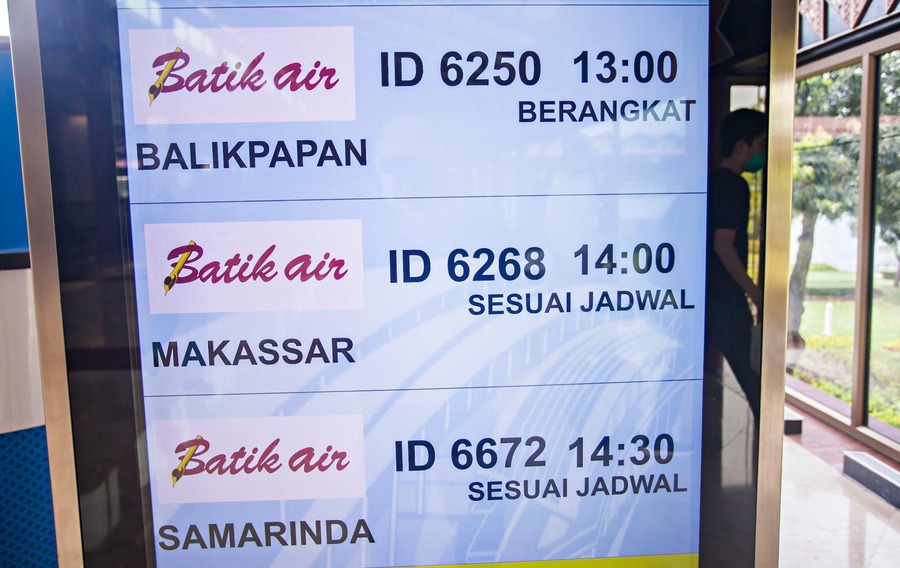}
\trajimglabel{Input image}
\vspace{4pt}
\includegraphics[width=\linewidth]{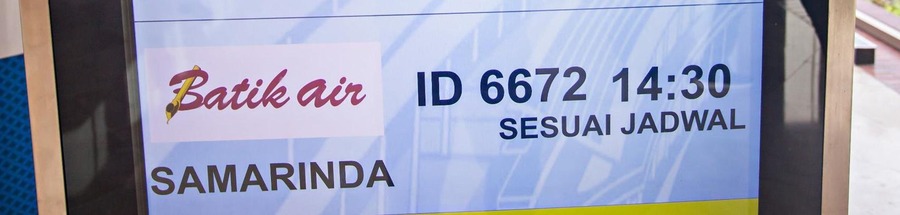}
\trajimglabel{Visual reference for supervision}
\end{minipage}
\hfill
\begin{minipage}[t]{0.57\textwidth}
\vspace{0pt}
\trajfield{Query}{What is the status of the Samarinda flight?}
\trajrole{Understanding}{The image shows a Batik Air flight display with rows for Balikpapan, Makassar, and Samarinda.}
\trajrole{Reasoning}{The relevant evidence is the row whose destination is Samarinda, so the model should focus on that row before answering.}
\trajrole{Visual}{A visual thought isolates the Samarinda row, including flight ID 6672, time 14:30, destination Samarinda, and its status field.}
\trajrole{Reasoning}{The isolated row states ``SESUAI JADWAL,'' which indicates that the flight is on schedule.}
\trajrole{Answer}{sesuai jadwal}
\end{minipage}
\end{trajcase}

\begin{trajcase}[borderline west={1.3pt}{0pt}{hard_color}, colframe=hard_color!35!black]{$p_{\mathrm{H}}$: Understanding $\rightarrow$ Reasoning $\rightarrow$ Hypothesis $\rightarrow$ Reasoning $\rightarrow$ Answer}
\begin{minipage}[t]{0.38\textwidth}
\vspace{0pt}
\centering
\includegraphics[width=0.92\linewidth]{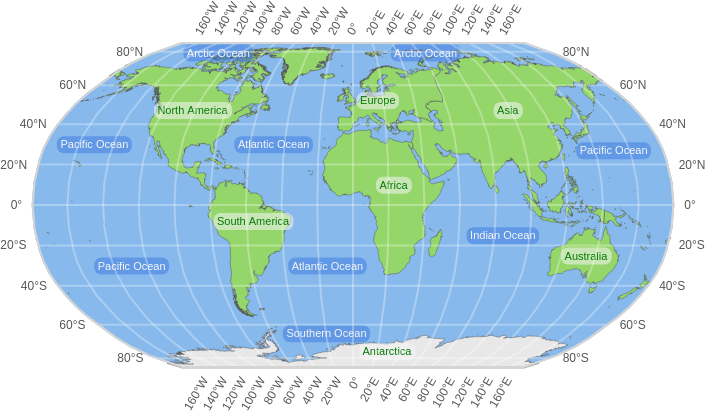}
\trajimglabel{Input image}
\vspace{3pt}
\begin{minipage}[t]{0.47\linewidth}
  \centering
  \includegraphics[width=\linewidth]{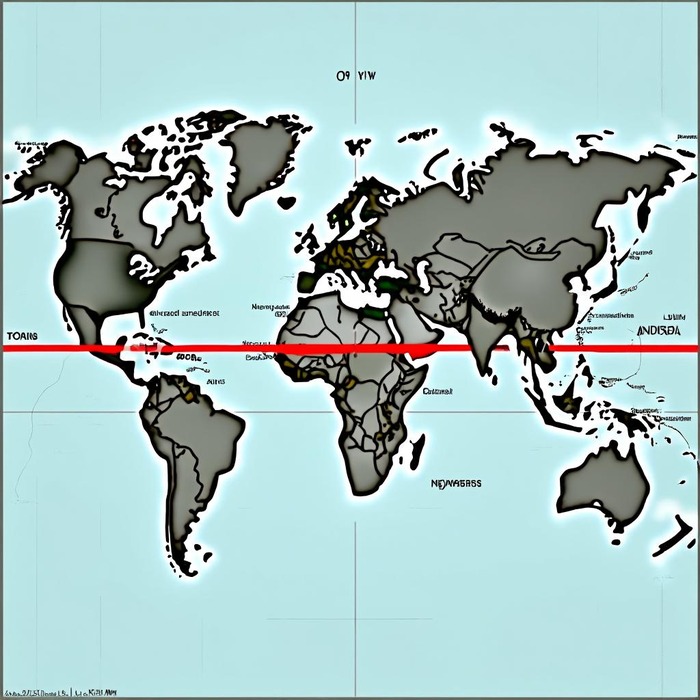}
  {\scriptsize Candidate 1\par}
\end{minipage}
\hfill
\begin{minipage}[t]{0.47\linewidth}
  \centering
  \includegraphics[width=\linewidth]{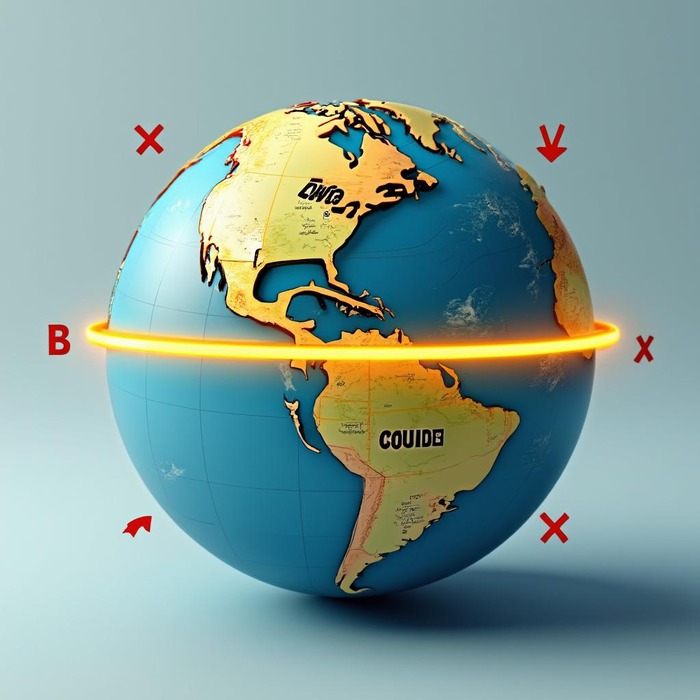}
  {\scriptsize Candidate 2\par}
\end{minipage}
\trajimglabel{Hypothesis references}
\end{minipage}
\hfill
\begin{minipage}[t]{0.57\textwidth}
\vspace{0pt}
\trajfield{Query}{Which of these continents does the equator intersect? Choices: (A) Europe, (B) Asia, (C) Antarctica.}
\trajrole{Understanding}{The image shows a world map with continents, oceans, and latitude lines, including the equatorial region.}
\trajrole{Reasoning}{The answer can be found by checking which candidate continent crosses the equator at 0 degrees latitude.}
\trajrole{Hypothesis}{Candidate references mark the equator on a world map or globe, making the possible land intersections easier to compare.}
\trajrole{Reasoning}{The equator does not pass through Europe or Antarctica. It crosses island regions of Asia, so the matching choice is Asia.}
\trajrole{Answer}{Asia}
\end{minipage}
\end{trajcase}

\phantomsection
\section*{Limitations}
\label{app:limitations}

The main limitation of our current system is path selection. Our results show a large remaining gap between the deployable planner and oracle path selection, which means that the proposed path space contains substantial unused potential. Closing this gap requires a planner that is both more accurate at the instance level and more robust across domains. This is challenging because planner supervision is expensive: to know which path works for an input, we must run multiple candidate paths and compare their outcomes. Such supervision is much more costly than ordinary single-trajectory training, especially for paths that involve visual-thought construction or hypothesis exploration.

The same issue also limits cross-dataset robustness. Different benchmarks have markedly different domain and query-form distributions, and our query-form buckets are calibrated on auxiliary data. As a result, the method implicitly assumes that the target test distribution is reasonably aligned with the calibration distribution. Dataset-adapted results suggest that performance can improve when the target distribution is known, but this weakens the goal of a single deployable planner. Future work should therefore focus on cheaper ways to collect path-outcome supervision and on planner objectives that generalize better under domain shift.

\phantomsection
\section*{Broader Impact}
\label{app:broader_impact}

This work studies how unified multimodal models coordinate understanding and generation. Potential positive impacts include more efficient multimodal reasoning, lower inference cost when a short path is sufficient, and more interpretable model behavior because the selected coordination path exposes when the model uses understanding, reasoning, visual thought, or hypothesis comparison. These properties may help developers analyze failures and build systems that spend computation more selectively.

The same capability can also create risks. Improvements in multimodal reasoning and generation can make synthetic visual content easier to produce or adapt, which may support disinformation or other deceptive uses if deployed without safeguards. Incorrect path selection can also make a system appear confident while using an inappropriate reasoning process, which is concerning in domains where visual answers affect user decisions. Because our planner uses query-form buckets calibrated on auxiliary data, performance may vary across domains, languages, or user groups whose inputs differ from the calibration distribution.

Mitigations include evaluating path selection across diverse domains and languages, reporting planner uncertainty or path traces to support auditing, and applying standard safeguards for generative systems such as content provenance, watermarking, misuse monitoring, and access controls when models are released. The system should not be used as an autonomous decision-maker in high-stakes settings without task-specific validation and human oversight.

\clearpage
\section*{NeurIPS Paper Checklist}

\begin{enumerate}

    \item {\bf Claims}
    \item[] Question: Do the main claims made in the abstract and introduction accurately reflect the paper's contributions and scope?
    \item[] Answer: \answerYes{} 
    \item[] Justification: The claims are stated in the abstract and introduction, supported in Sec.~4, and scoped by Appendix~\ref{app:limitations}.

\item {\bf Limitations}
    \item[] Question: Does the paper discuss the limitations of the work performed by the authors?
    \item[] Answer: \answerYes{} 
    \item[] Justification: The main limitations are discussed in Appendix~\ref{app:limitations}.

    \item {\bf Theory assumptions and proofs}
    \item[] Question: For each theoretical result, does the paper provide the full set of assumptions and a complete (and correct) proof?
    \item[] Answer: \answerNA{} 
    \item[] Justification: The paper does not present theoretical theorems or proof claims.

    \item {\bf Experimental result reproducibility}
    \item[] Question: Does the paper fully disclose all the information needed to reproduce the main experimental results of the paper to the extent that it affects the main claims and/or conclusions of the paper (regardless of whether the code and data are provided or not)?
    \item[] Answer: \answerYes{} 
    \item[] Justification: Reproducibility details are provided in Sec.~\ref{sec:experimental_setup} and Appendices~\ref{app:training_data_details}, \ref{app:planner_configuration}, \ref{app:executor_training_details}, and \ref{app:benchmark_metric_details}.

\item {\bf Open access to data and code}
    \item[] Question: Does the paper provide open access to the data and code, with sufficient instructions to faithfully reproduce the main experimental results, as described in supplemental material?
    \item[] Answer: \answerYes{} 
    \item[] Justification: Code and processed trajectories are released at the URL in the abstract.

\item {\bf Experimental setting/details}
    \item[] Question: Does the paper specify all the training and test details (e.g., data splits, hyperparameters, how they were chosen, type of optimizer) necessary to understand the results?
    \item[] Answer: \answerYes{} 
    \item[] Justification: Settings are described in Sec.~\ref{sec:experimental_setup} and Appendices~\ref{app:executor_training_details}, \ref{app:executor_data_stats}, and \ref{app:planner_configuration}.

\item {\bf Experiment statistical significance}
    \item[] Question: Does the paper report error bars suitably and correctly defined or other appropriate information about the statistical significance of the experiments?
    \item[] Answer: \answerNo{} 
    \item[] Justification: Error bars are omitted because repeating the full multi-path executor and planner pipeline is computationally expensive.

\item {\bf Experiments compute resources}
    \item[] Question: For each experiment, does the paper provide sufficient information on the computer resources (type of compute workers, memory, time of execution) needed to reproduce the experiments?
    \item[] Answer: \answerYes{} 
    \item[] Justification: Compute resources are reported in Appendix~\ref{app:executor_training_details}.
    
\item {\bf Code of ethics}
    \item[] Question: Does the research conducted in the paper conform, in every respect, with the NeurIPS Code of Ethics \url{https://neurips.cc/public/EthicsGuidelines}?
    \item[] Answer: \answerYes{} 
    \item[] Justification: We confirm that the work follows the NeurIPS Code of Ethics.

\item {\bf Broader impacts}
    \item[] Question: Does the paper discuss both potential positive societal impacts and negative societal impacts of the work performed?
    \item[] Answer: \answerYes{} 
    \item[] Justification: Broader impacts are discussed in Appendix~\ref{app:broader_impact}.
    
\item {\bf Safeguards}
    \item[] Question: Does the paper describe safeguards that have been put in place for responsible release of data or models that have a high risk for misuse (e.g., pre-trained language models, image generators, or scraped datasets)?
    \item[] Answer: \answerNA{} 
    \item[] Justification: The submission does not release a new high-risk pretrained model, image generator, or scraped image dataset.

\item {\bf Licenses for existing assets}
    \item[] Question: Are the creators or original owners of assets (e.g., code, data, models), used in the paper, properly credited and are the license and terms of use explicitly mentioned and properly respected?
    \item[] Answer: \answerYes{} 
    \item[] Justification: Existing datasets, benchmarks, and models are credited.

\item {\bf New assets}
    \item[] Question: Are new assets introduced in the paper well documented and is the documentation provided alongside the assets?
    \item[] Answer: \answerYes{} 
    \item[] Justification: The processed trajectories are documented in Appendix~\ref{app:training_data_details} and will be released.

\item {\bf Crowdsourcing and research with human subjects}
    \item[] Question: For crowdsourcing experiments and research with human subjects, does the paper include the full text of instructions given to participants and screenshots, if applicable, as well as details about compensation (if any)? 
    \item[] Answer: \answerNA{} 
    \item[] Justification: The work does not involve crowdsourcing or human-subject experiments.

\item {\bf Institutional review board (IRB) approvals or equivalent for research with human subjects}
    \item[] Question: Does the paper describe potential risks incurred by study participants, whether such risks were disclosed to the subjects, and whether Institutional Review Board (IRB) approvals (or an equivalent approval/review based on the requirements of your country or institution) were obtained?
    \item[] Answer: \answerNA{} 
    \item[] Justification: The work does not involve crowdsourcing or human-subject experiments, so IRB approval is not applicable.

\item {\bf Declaration of LLM usage}
    \item[] Question: Does the paper describe the usage of LLMs if it is an important, original, or non-standard component of the core methods in this research? Note that if the LLM is used only for writing, editing, or formatting purposes and does \emph{not} impact the core methodology, scientific rigor, or originality of the research, declaration is not required.
    \item[] Answer: \answerYes{} 
    \item[] Justification: The paper describes LLM use in Sec.~\ref{sec:experimental_setup} and Appendix~\ref{app:training_data_details}. 

\end{enumerate}

\end{document}